\documentclass[prb,preprint]{revtex4}

\pdfoutput=1

\usepackage{graphicx}
\begin{document}


\title{Stripes in oxygen-enriched cuprates}
\medskip 

\date{December 2, 2020} \bigskip

\author{Manfred Bucher \\}
\affiliation{\text{\textnormal{Physics Department, California State University,}} \textnormal{Fresno,}
\textnormal{Fresno, California 93740-8031} \\}

\begin{abstract}
Charge-order stripes of different types occur when copper oxides are  doped with either heterovalent metal, like $La_{2-x}Sr_xCuO_4$, or oxygen, like $YBa_2Cu_3O_{6+y}$. The difference shows up in the doping dependence of their incommensurability: $q_c(x) \propto \sqrt{x-\check{p}}$ but $q_c(y) \approx 0.3$. The square-root dependence in the former compound family results from Coulomb repulsion between doped holes (or electrons), residing pairwise in lattice-site $O$ (or $Cu$) atoms of the $CuO_2$ planes. The almost constant $q_c(y)$ value in the second family results from the aggregation of ozone-like molecules, formed from $O^{2-}$ ions of the host with embedded oxygen atoms, $\mathring{O}$, at interstitial sites in the $CuO_2$ planes.
The magnetic moments, $\mathbf{m}(O)$, of the lattice-defect $O$ atoms in the first family arrange antiferromagnetically, which gives rise to accompanying magnetization stripes of incommensurability $q_m(x) = \frac{1}{2} q_c(x)$. The ozone complex has a vanishing magnetic moment, $\mathbf{m}=0$, which explains the absence of accompanying magnetization stripes in the second family. Embedding excess oxygen as $\mathring{O}$ atoms in $CuO_2$ planes is likewise assumed for $HgBa_2CuO_{4+\delta}$ and oxygen-enriched bismuth cuprates.
A combination of characteristics from both families is present in oxygen-enriched $La_2CuO_{4+\delta}$. 
The validity of determining the hole density in oxygen-enriched cuprates with the universal-dome method is independently confirmed.
Besides causing different types of stripes, the two types of lattice-defect oxygen may also cause different types of superconductivity. This could explain the much higher $T_{c,max}$ in oxygen-enriched than $Sr$-doped cuprates, as well as the cusped cooling-curves of X-ray intensity diffracted by stripes in the former family.
\end{abstract}

\maketitle

\pagebreak

\section{INTRODUCTION}
Two types of charge-order stripes are observed in copper oxides of  high transition temperature (``high $T_c$'') superconductivity. The distinction is characterized by the doping dependence of their incommensurability and is determined by the doping species---whether doped with heterovalent metal or oxygen.
Doping with heterovalent metal results in hole doping when 
divalent alkaline earth ions, $Ae^{2+} = Sr^{2+}, Ba^{2+}$, partially substitute trivalent lanthanide ions, $Ln^{3+}=La^{3+},Nd^{3+}$, but in electron doping by substitution with tetravalent cerium, $Ce^{4+}$. Examples are $La_{2-x}Ae_xCuO_4$ and $Nd_{2-x}Ce_xCuO_4$. All compounds of this family have the `214' crystal structure (of $T$ or $T'$ type).
Examples of oxygen doping (or enrichment) are
$YBa_2Cu_3O_{6+y}$, $HgBa_2CuO_{4+\delta}$, $Bi_2Sr_{2-x}La_xCuO_{6+\delta}$ 
and $Bi_2Sr_2CaCu_2O_{8+\delta}$.  
Common to both families is the presence of $CuO_2$ planes where the stripes reside (and where  superconductivity occurs).
However, the doping dependence of their stripe incommensurability is distinctly different: 
For the family of the `214' compounds it is given, in reciprocal lattice units (r.l.u.), by
\begin{equation}
q_c(x)  =  \frac{\Omega^{\pm}}{2}\sqrt {x - \check{p}} \;,  \;\;\;\; x\le\hat{x}\; ,
\end{equation}
\noindent for $Ae$ doping $x$ up to a ``watershed'' value $\hat{x}$.\cite{1}
The stripe-orientation factor is $\Omega^{+}=\sqrt{2}$ for $x > 0.056$ when stripes are parallel to the $a$ or $b$ axis, but $\Omega^{-} = 1$ for $x < x_6$ when stripes are diagonal. The offset value under the radical is $\check{p} \le 0.02$ for the hole-doped `214' compounds.
A qualitative change of the incommensurability occurs at a watershed concentration of the dopant, $\hat{x}$, which depends on the species of doping and co-doping. It shows up as \emph{kinks} in the $q_c(x)$ profile at $\hat{x}$ (see Fig. 1, left), where the square-root curve from Eq.(1) levels off to constant plateaus,
\begin{equation}
    q_c(x) = \frac{\sqrt{2}}{2} \sqrt{\hat{x} - \check{p}} \;,\;\;\;\; x > \hat{x} \;.
\end{equation}
The charge-order stripes in the `214' family are accompanied by magnetization stripes of incommensurability
\begin{equation}
    q_m(x) = \frac{1}{2} q_c(x) \;.
\end{equation}

In stark contrast, the incommensurability of charge-order stripes in $YBa_2Cu_3O_{6+y}$---and similarly in the oxygen-enriched $Hg$ and $Bi$
compounds---depends only weakly on oxygen doping,
$q_c(y) \approx 0.3$, {\it decreasing} slightly with super-oxygenation (see Fig. 1). Noticeably, those stripes are \emph{not} accompanied (at the same doping $y$) by magnetization
stripes. A  re-

\includegraphics[width=5.6in]{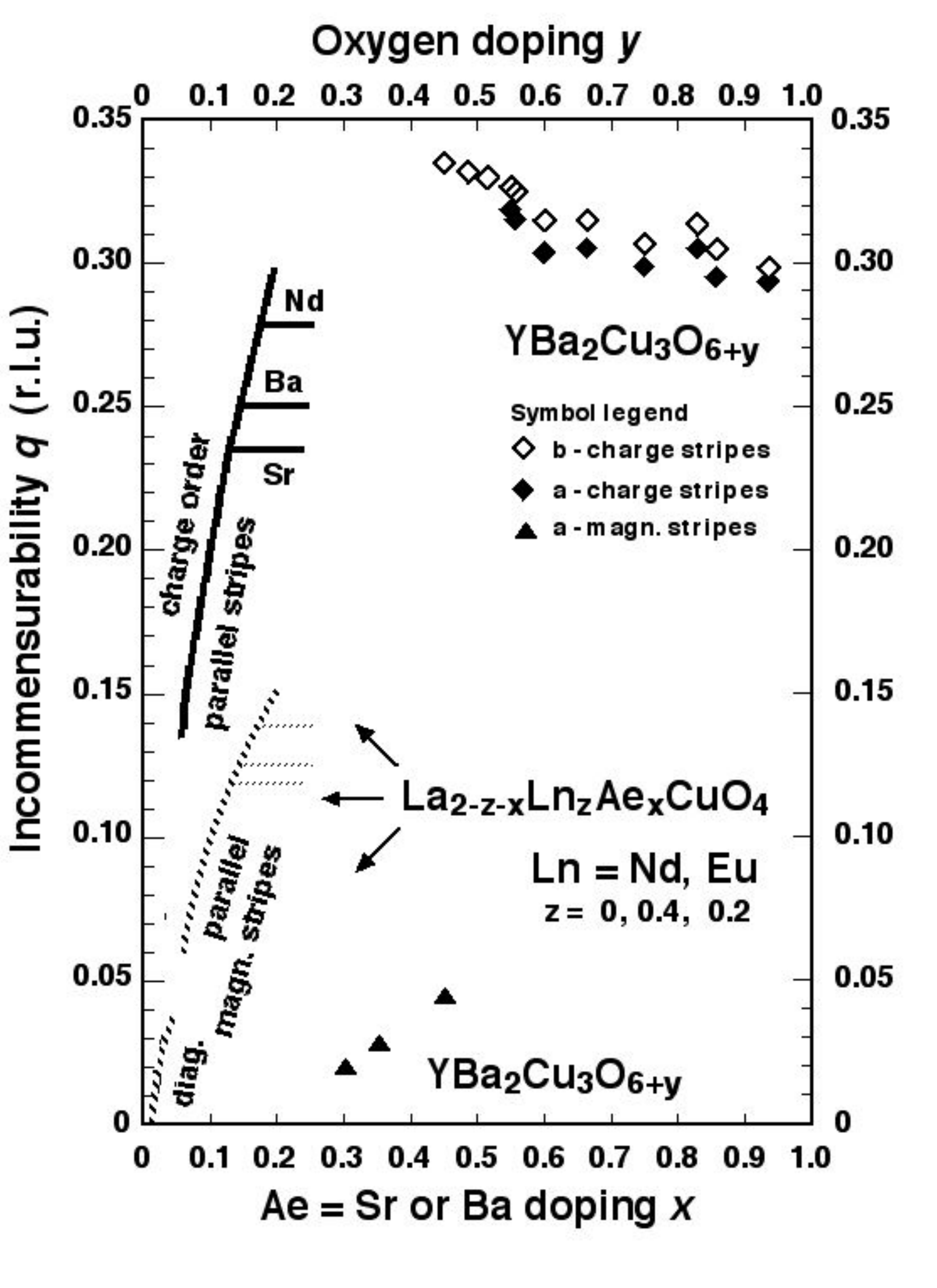}

\footnotesize 

\noindent FIG. 1. Incommensurability of charge-order stripes (solid line) and magnetization stripes (hatched line) in $La_{2-z-x}Ln_zAe_{x}CuO_{4}$ of the `214' family due to $Ae$ doping with $Ae = Sr$ or $Ba$ (curves on the left) and in oxygen-doped $YBa_2Cu_3O_{6+y}$ (data from Refs. 2-17) near the top and bottom. The curves are graphs of Eqs. (1-3).
The discontinuity at $x = 0.056$ is caused by a change of stripe orientation from diagonal to parallel, relative to the $Cu$-$O$ bonds.
 
\pagebreak \normalsize 

\noindent sult of insufficient annealing,\cite{6} magnetization stripes of square-root doping dependence, akin to Eq. (1), have been observed in $YBa_2Cu_3O_{6+y}$ in the middle range of oxygen doping,\cite{5} $0.30 \le y \le 0.45$, preceding the doping range of charge-order stripes, $0.45 \le y \le 0.92$.
Since the incommensurability is much better understood in the $Ae$-doped `214' compounds\cite{1} than in the oxygen-doped cuprates, knowledge about the former can be used to make inferences about the observed similarities and differences with stripes in the latter family. This is the aim of the present paper. To this end the nature of stripes in the `214' family is briefly reviewed.

\section{STRIPES IN LANTHANUM CUPRATES DOPED WITH ALKALINE-EARTH}

The unit cell of \emph{pristine} ${La_2CuO_4}$ has a central $CuO_2$ plane, sandwiched by $LaO$ planes (see Fig. 2, left). Consider stepwise ionization, where brackets indicate electron localization at atoms, both within the planes and by transfer from the $LaO$ planes to the $CuO_2$ plane:
\bigskip 
\newline
\noindent $LaO \;\;:\; La^{3+} + 3e^- + \;\;O \rightarrow  La^{3+} + [2e^- + O]\; + 
\downarrow \overline {e^-\;} | \rightarrow  La^{3+} + O^{2-}$
\newline
$CuO_2 :\; Cu^{2+} + 2e^- + 2O \rightarrow  Cu^{2+} + [2e^- + O] + \;\;\;O \;\;\;\rightarrow  O^{2-} \;+  Cu^{2+} + O^{2-}$
\newline
$LaO \;\;:\; La^{3+} + 3e^- + \;\;O \rightarrow  La^{3+} + [2e^- + O]\; + 
\uparrow \underline {e^-\;} | \rightarrow  La^{3+} + O^{2-}$
\bigskip

\noindent In the simplest case of {\it doping},
$Ae$ substitutes,  in some cells, $La$ in both sandwiching planes: 

\bigskip 
\noindent $AeO \;\;:\; Ae^{2+} + 2e^- + \;\;O \rightarrow  Ae^{2+} + [2e^- + O]\; \;\;\;\;\;\;\;\;\;\;\;\; \;\rightarrow  Ae^{2+} + O^{2-}$
\newline
$CuO_2 :\; Cu^{2+} + 2e^- + 2O \rightarrow  Cu^{2+} + [2e^- + O] + \;\;\;O \;\;\;\rightarrow  O^{2-} \;\;+  Cu^{2+} + \mathbf{\tilde{O}}$
\newline
$AeO \;\;:\; Ae^{2+} + 2e^- + \;\;O \rightarrow  Ae^{2+} + [2e^- + O]\; \;\;\;\;\;\;\;\;\;\;\;\;\; \rightarrow  Ae^{2+} + O^{2-}$
\bigskip

\noindent The lack of electron transfer from the sandwiching planes to the $CuO_2$ plane leaves some oxygen atoms \textit{neutral} (marked bold above and below). Compared to the host crystal, they can be regarded as housing the holes (pairwise). Up to a density $\check{p} \le  0.02$, such holes are \textit{itinerant}, enabling $\mathbf{\tilde{O}}$ atoms to skirmish long-range antiferromagnetism (3D-AFM). The remaining lack of electron transfer leaves more oxygen atoms \textit{stationary} at lattice sites, $\mathbf{O}$. They give rise to static stripes (to be explained instantly).

\bigskip 
\noindent $AeO \;\;:\; Ae^{2+} + 2e^- + \;\;O \rightarrow  Ae^{2+} + [2e^- + O]\; \;\;\;\;\;\;\;\;\;\;\;\; \;\rightarrow  Ae^{2+} + O^{2-}$
\newline
$CuO_2 :\; Cu^{2+} + 2e^- + 2O \rightarrow  Cu^{2+} + [2e^- + O] + \;\;\;O \;\;\;\rightarrow  O^{2-} \;\;+  Cu^{2+} + \mathbf{O}$
\newline
$AeO \;\;:\; Ae^{2+} + 2e^- + \;\;O \rightarrow  Ae^{2+} + [2e^- + O]\; \;\;\;\;\;\;\;\;\;\;\;\;\; \rightarrow  Ae^{2+} + O^{2-}$
\bigskip

\noindent Relative to the host crystal, both the skirmishing and stationary oxygen atoms, $\tilde{O}$ and $O$, 
have a \emph{lattice-defect charge}, $Q^{latt}_{def} =+2|e|$, holding two elementary charges each (see Appendix A). 
In this sense, the doping of $La_2CuO_4$ with $Ae$ is often called ``hole doping'' (more accurately, doping the $CuO_2$ planes with holes). 
Both the $\tilde{O}$ and $O$ atoms have a
finite magnetic moment, $\mathbf{m}(\tilde{O}) =\mathbf{m}(O) \ne 0 $, due to their spin quantum number $S = 2\times \frac{1}{2} = 1$ according to the spin configuration $[\uparrow\downarrow]$ $[\uparrow]$ $[\uparrow]$ of their $2p^4$ subshell (Hund's rule of maximal multiplicity). 
The moments of the skirmishing oxygen atoms, $\mathbf{m}(\tilde{O})$---itinerant \textit{via} anion lattice sites---upset the 3D-AFM the host [from $\mathbf{m}(Cu^{2+})$ moments] and cause its collapse at hole density $\check{p}=0.02$ (more in Sect. V).

The superlattice, formed by the $O$ ions, gives rise to charge-order stripes and magnetization stripes with a square-root dependence of the incommensurability on doping, Eqs. (1, 3). 
The square-root dependence of $q_c(x)$ results from the spreading of the double holes by Coulomb repulsion to the farthest available separations. 
Thus a rising square-root profile of stripe incommmensurability
signifies an underlying superlattice of lattice-defect charges.
The charge-order and magnetization stripes are coupled, Eq. (3), because the constituting charges and magnetic moments reside at the \emph{same sites}---the lattice-defect $O$ atoms. 
It is the antiparallel orientation of neighboring magnetic moments $\mathbf{m}(O)$---in the $O$ superlattice---that doubles the periodic length of the magnetization stripes. 
Increasing density of the doped holes, housing pairwise at lattice-defect $O$ atoms in the $CuO_2$ planes, raises their 
Coulomb repulsion energy. When doping exceeds a watershed value, $x > \hat{x}$, additional holes overflow to the $LaO$ planes\cite{1} where they also reside pairwise in $O$ atoms. This leaves charge-order stripes of \emph{constant} $q_c$ in the $CuO_2$ planes, Eq. (2).

\section{STRIPES IN $\mathbf{n}$-DOPED $\mathbf{Nd_{2-x}Ce_xCuO_{4+\delta}}$ AND OXYGENATED $\mathbf{La_2CuO_{4+\delta}}$}    
    
The unit cell of \emph{pristine} ${Nd_2CuO_4}$ has a central $CuO_2$ plane, sandwiched by $NdO$ planes, analogous to ${La_2CuO_4}$ in Fig. 2. Consider again crystal formation by stepwise ionization, where brackets indicate electron localization at atoms, both within the planes and by transfer from the $NdO$ planes to the $CuO_2$ plane:
\bigskip 

\noindent $NdO \;\;:\; Nd^{3+} + 3e^- + \;O \rightarrow  Nd^{3+} + [2e^- + O]\; + 
\downarrow \overline {e^-\;} | \rightarrow  Nd^{3+} + O^{2-}$
\newline
$CuO_2 :\; Cu^{2+} + 2e^- + 2O \rightarrow  Cu^{2+} + [2e^- + O] \;+ \;\;\;O \;\;\;\rightarrow \; O^{2-} \;\;+  Cu^{2+} + O^{2-}$
\newline
$NdO \;\;:\; Nd^{3+} + 3e^- + \;O \rightarrow  Nd^{3+} + [2e^- + O]\; + 
\uparrow \underline {e^-\;} | \rightarrow  Nd^{3+} + O^{2-}$
\bigskip

\noindent In the simplest case of {\it doping},
$Ce$ substitutes, in some cells, $Nd$ in both sandwiching planes: 

\bigskip 

\noindent $CeO \;\;:\; Ce^{4+} + 4e^- + \;\;O \rightarrow | \overline {e^-} \downarrow + \; Ce^{4+} + [2e^- + O]\; + 
\downarrow \overline {e^-\;} | \rightarrow  Ce^{4+} + O^{2-}$
\newline
$CuO_2 :\; Cu^{2+} + 2e^- + 2O \rightarrow  Cu^{2+} + 
\;\;\;\;\;\;\;\;\;\;\;\;[2e^- + O] + \;\;\;O \;\;\;\rightarrow    \mathbf{Cu} \;+ \;2O^{2-}$
\newline
$CeO \;\;:\; Ce^{4+} + 4e^- + \;\;O \rightarrow | \underline {e^-} \uparrow +\; Ce^{4+} + [2e^- + O]\; + 
\uparrow \underline {e^-\;} | \rightarrow  Ce^{4+} + O^{2-}$
\bigskip

\noindent Doping lanthanide-based cuprates with cerium, $Ln_{2-x}Ce_xCuO_{4+\delta} \;  (Ln = Nd, Pr, La$, $Sm, Eu)$, partially substitutes $Ln^{3+}$  by $Ce^{4+}$, resulting in \emph{electron doping} of the $CuO_2$ planes. 
The doped electrons reside pairwise in lattice-site $\mathbf{Cu}$ atoms.\cite{1}
For reasons of stability, the crystals need to be grown with excess oxygen, $O_\delta$, to be eliminated in post-growth annealing.\cite{18} The incommensurability of charge-order stripes (for $x > x_6$) is given by Eq. (1) with $\check{p} = 2\delta$.\cite{1} As Eq. (1) is based on Coulomb repulsion of like charges in the $CuO_2$ planes, its success for stripes in electron-doped `214' compounds (of $T'$ structure) implies that the excess oxygen must reside interstitially as oxygen \textit{ions}, $O_\delta^{2-}$, in or between the $LnO$ planes (see Fig. 2).\cite{1} It results in a \emph{hole}-doping contribution to the $CuO_2$ planes that correspondingly reduces their electron-doping from $Ce$.
Onset of superconductivity in nominal $Nd_{2-x}Ce_xCuO_4$ is observed at $Ce$ doping $x\simeq 0.13$.\cite{18} However, recent experiments have shown onset of superconductivity at much lower $x$ when additional electron doping is accomplished by \emph{oxygen deficiency}, $Nd_{2-x}Ce_xCuO_{4-\delta}$.\cite{19}

Concerning the Mott insulator $La_2CuO_4$, superconductivity is achieved not only by doping with heterovalent metal, $La_{2-x}M_xCuO_4 \;(M = Sr, Ba, Ce)$, but also by {\it oxygen enrichment}, $La_2CuO_{4+\delta}$. 
However, there are notable differences.
Investigations with neutron scattering and X-ray spectroscopy have shown that $La_2CuO_{4+\delta}$ undergoes a phase separation into domains that are rich or poor in interstitial oxygen, $O_i=O_\delta$.\cite{20,21,22,23,24} This is in contrast to homogeneous doping in the $La_{2-x}M_xCuO_4$ compounds. Superconductivity in $La_2CuO_{4+\delta}$ is granular, occurring in the $O_i$-rich domains, with only three discrete values of the transition temperature ($T_c = 17, 32, 42$ K) instead of a continuous superconducting $T_c$ dome.\cite{24,25,28}

A structural analysis from neutron scattering\cite{20} assigns interstitial oxygen in the $O_i$-rich domains to the $(\frac{1}{4},\frac{1}{4},-\frac{1}{6})$ positions between staggered $LaO$ planes, called $O(4)$ (see Fig. 2).
Occupying $O(4)$ positions between successive $LaO$ double planes, interstitial oxygen ions align  parallel to the crystal's space diagonal, forming {\it commensurate} $O_i^{2-}$ stripes of commensurability  $(q_i^b, q_i^c)^ {ortho} = (0.25, 0.5)$ in orthorhombic r.l.u.\cite{22,23,24}

Residence of \emph{ionized} excess oxygen at the $O(4)$ positions causes hole doping of the $CuO_2$ planes, where the holes reside pairwise in $O$ atoms, as in $La_{2-x}Ae_xCuO_4$.  Their magnetic moments, $\mathbf{m}(O)\ne0$, give rise to magnetization stripes, as has been observed  with neutron scattering.\cite{22,27} 
[Results from the two neutron investigations differ slightly, possibly as a result of different oxygen-enrichment procedures (electro-chemical, aqueous bath). Here we average the values from both investigations.]
The magnetization stripes are oriented closely along the $Cu$-$O$ bonds
(rotated off by $\sim 5^{\circ}$),\cite{22,27} with equal fractions along the tetragonal $a^T$ and $b^T$ axes.\cite{28}
 Their incommensurability, converted to tetragonal r.l.u., is $q_m \simeq 0.115$. 
From the $q_c(x) = 2q_m(x)$ pattern of stripes in $La_{2-x}Ae_xCuO_4$, one would expect charge-order stripes in $La_2CuO_{4+\delta}$ parallel to the $Cu$-$O$ bonds, with $q_c \simeq 2 \times 0.115 = 0.23$. Instead, charge-order stripes of $q_c = 0.15$ are observed (converted to quasi-tetragonal r.l.u. from $q_c^{b, ortho} =  0.21$), \emph{diagonal} to the $Cu$-$O$ bonds.\cite{21,24,26} How can that be understood?
 
 In $Ae$-doped $La_{2-x}Ae_xCuO_4$, an incommensurability $q_m = 0.115$  results, by Eqs. (3, 1), from a hole density $p=x=0.12$ in the $CuO_2$ planes. Since the same $q_m=0.115$ is observed in oxygen-enriched $La_2CuO_{4+\delta}$, a hole density of $p=0.12$ can also be assumed in that compound. The excess oxygen in $La_2CuO_{4+\delta}$ has been determined by gravitometry as $\delta = 0.12$.\cite{22} If all excess oxygen were embedded as $O_i^{2-}$ ions at $O(4)$ sites, this would result in a hole density $p = 2\delta = 0.24$.
 The finding that magnetization stripes in $La_2CuO_{4+\delta}$ are caused by only \emph{half} of the oxygen enrichment leaves open the possibility that the other half involves an embedding  mechanism that is non-magnetic but modifies the charge order. We will return to this possibility in Sect. XII.

\section{CRYSTAL STRUCTURE AND DOPING}

Before further comparison of stripes in $La_{2-x}Ae_xCuO_4$ and $YBa_2Cu_3O_{6+y}$, we need to get more familiar with the crystal structure of the compounds. Figure 2 shows a double unit cell of pristine $La_2CuO_4$ and a unit cell of $YBa_2Cu_3O_6$. The former crystal has the simpler structure, with $LaO$ planes sandwiching the $CuO_2$ planes. However, its unit cells are vertically \emph{staggered} with displacements by half the planar lattice constants, ($\frac{a_0}{2}, \frac{b_0}{2}$). 
Pristine $La_2CuO_4$ is orthorhombic with lattice constants $a_0 < b_0 < c_0$. Partial replacement of $La^{3+}$ by the larger-sized $Ae^{2+}= Sr^{2+}, Ba^{2+}$ ions creates internal stress from lattice mismatch. The stress is relieved by a low-temperature phase transition to a tetragonal structure, $a_0 = b_0 < c_0$, occurring near $x = 0.125$ when $Ba$-doped, but at $x = 0.21$ when $Sr$-doped.

In $YBa_2Cu_3O_6$, two $CuO_2$ planes sandwich the central $Y$ plane. The $CuO_2$-$Y$-$CuO_2$  
\medskip 

\includegraphics[width=6in]{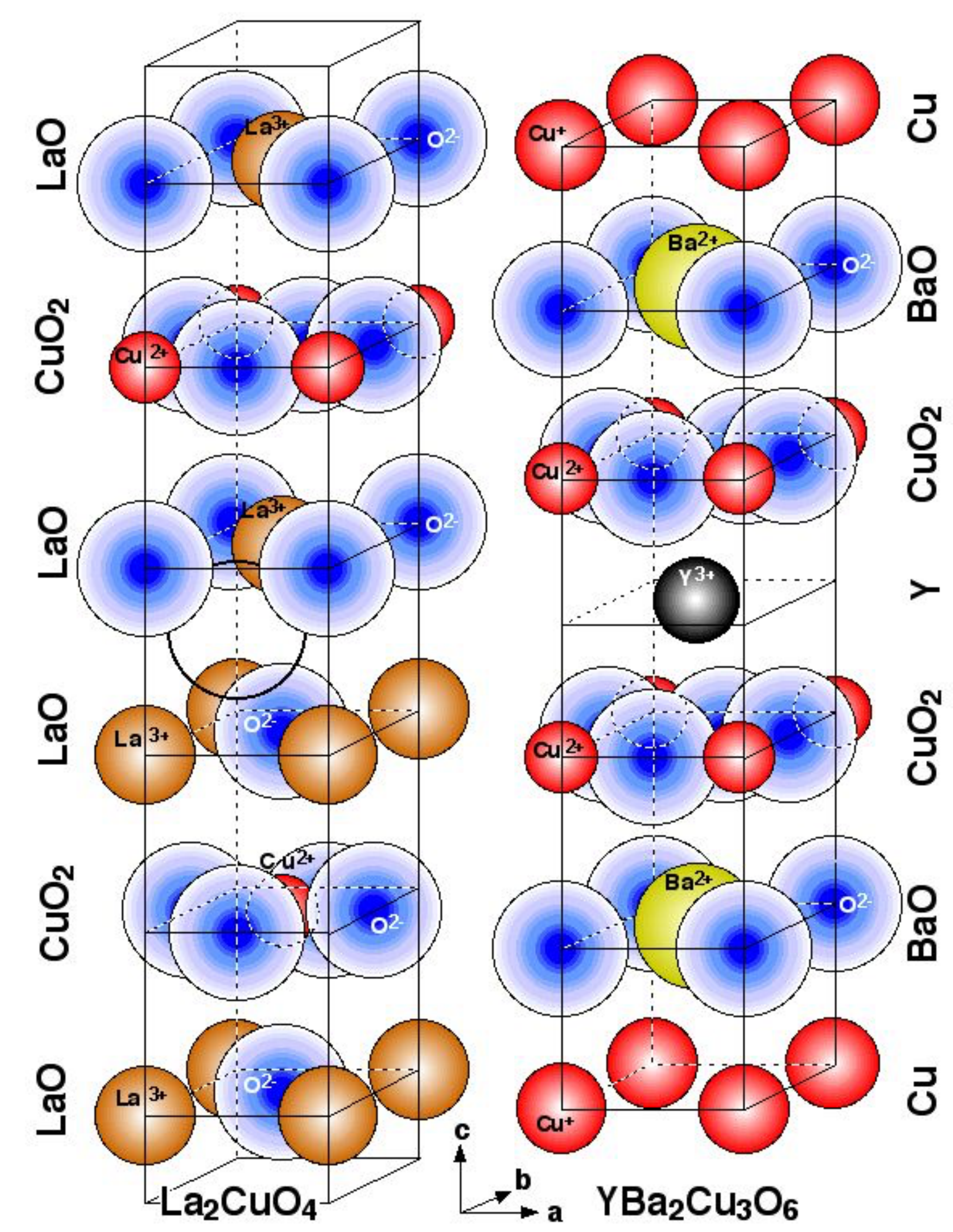} \footnotesize 

\noindent FIG. 2.  Two unit cells of $La_2CuO_4$ (left) and single unit cell of $YBa_2Cu_3O_6$ (right), vertially exploded.  Ion planes are noted on the sides. The black ring indicates an $O(4)$ site for an interstitial  $O^{2-}$ ion in $La_2CuO_4$. 
 \normalsize 

\noindent layer in turn is sandwiched by $BaO$- and terminal $Cu$-planes. Contrary to the `214' compounds,  $YBa_2Cu_3O_6$ is tetragonal but becomes orthorhombic upon oxygen doping at $y \approx 0.3$. Notice that by stoichiometry and charge balance, $Y^{3+}Ba^{2+}_2(Cu^{2+}_2Cu^+)O^{2-}_6$, the copper ions are in different ionization states, with $Cu^{2+}$ ions in the $CuO_2$ planes but $Cu^+$ ions in the $Cu$ planes (see Fig. 2). There are four possibilities how a doped oxygen atom can be embedded, labelled \#1 to \#4. For the sake of clarity they are introduced with the idealized assumption of a ``0.5-watershed,'' meaning that processes \#1 and \#2 occur only for oxygen doping $y \le 0.5$ and processes \#3 and \#4 only for $y>0.5$. The restriction, which captures most of the compound's properties, will be relaxed later to account for some additional aspects.

\bigskip \noindent \textbf{\#1.} A doped oxygen atom can be incorporated into the $Cu$ plane between two $Cu^+$ ions, along the $b$-direction, contributing to a so-called $CuO$ \textit{chain} (see Fig. 3). In this case the oxygen atom takes two electrons from two neighboring copper ions in the $Cu$ plane, denoted as $Cu^+(1)$. It results in higher ionization states of all three entities,
\begin{equation}
    O + \; 2 \; Cu^+(1)  \rightarrow \; O^{2-}(1) + \; 2 \; Cu^{2+}(1) \;, \;\;\;\;\; 0 < y \le 0.5 \; .
\end{equation}
The $O(1) \rightarrow O^{2-}(1)$ embedding process can occur only for $y\le0.5$ but is exhausted at $y=0.5$ when all former $Cu^+(1)$ ions have been ionized to $Cu^{2+}(1)$ ions. This can be seen by stoichiometry and charge balance of the resulting compound,
$Y^{3+}Ba^{2+}_2Cu^{2+}_3O^{2-}_{6.5}$.
It seems favorable, in terms of binding energy, that the doped oxygen aligns as $O^{2-}(1)$ ions in $CuO$ chains. For $y=0.5$ the configuration consists of alternate full and empty $O^{2-}(1)$ rows along the $b$-direction of the $Cu$ planes, called ``ortho-II oxygen ordered'' (see Fig. 3). Note that process \# 1 does not generate holes and thus \emph{not} give rise to charge-order stripes in the $CuO_2$ planes.

\bigskip \noindent \textbf{\#2.} The second possibility of embedding is a variation of \#1 that occurs as a result of imperfect crystal growth (here, insufficient annealing), but is absent in well-annealed crystals. Nevertheless, this (accidental) case provides some insight into stripe formation in $YBa_2Cu_3O_{6+y}$ and establishes a connection with stripes in the `214' compounds, $La_{2-x}Ae_xCuO_4$. With increased filling of the $CuO$ chains at oxygen doping $y < 0.5$, not every embedding $O$ atom may find two close $Cu^+$ ions to take up two electrons. In this case it can take one electron from one nearby $Cu^+$ ion and ``borrow'' one electron from an oxygen ion in the $CuO_2$ plane, denoted as $O^{2-}(2)$, whose outer electron is less bound than in an $O^{2-}(2)$ ion of the intermediate $BaO$ plane,\cite{1} 
\begin{equation}
    O +  \; Cu^+(1) \; + \;  O^{2-}(2)  \rightarrow \; O^{2-}(1) + \; Cu^{2+}(1)  \; + \;  O^-(2) \;, \;\;\;\;\;0 < y < 0.5 \; .
\end{equation}
The process leaves temporarily an unstable $O^-(2)$ ion the $CuO_2$ plane, quickly to combine 
\medskip
\includegraphics[width=6.5in]{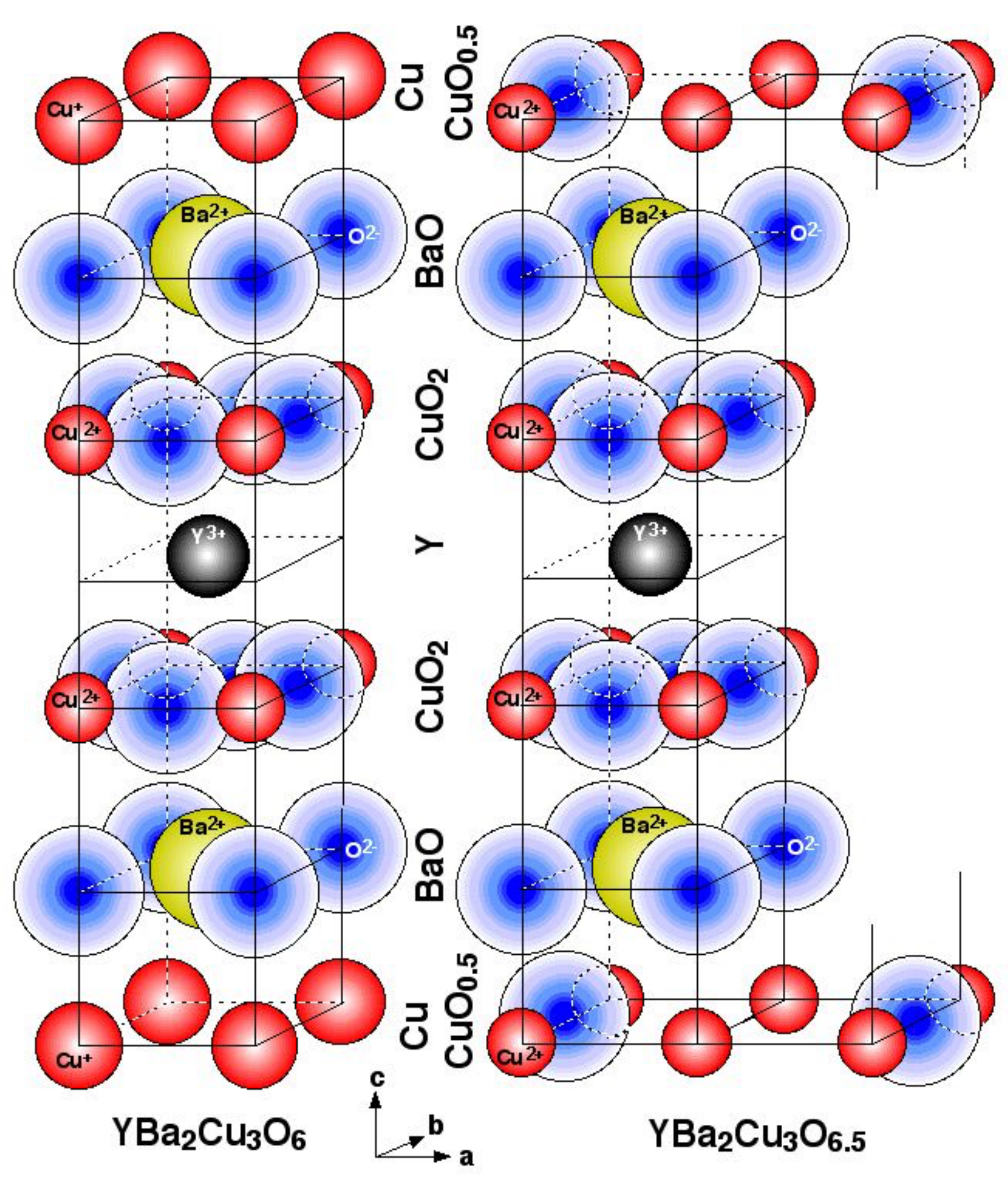}  \footnotesize 

\noindent FIG. 3.  Unit cell of $YBa_2Cu_3O_6$ (left) and of $YBa_2Cu_3O_{6.5}$ with ortho-II ordered $CuO$ chains (right). 

\pagebreak \normalsize 
\noindent with another one to form a lattice-site oxygen \emph{atom},
\begin{equation}
    2  \; O^-(2)  \rightarrow \; O^{2-}(2) + \; O(2) \;, \;\;\;\;\; 0 < y < 0.5  \; .
\end{equation}
The assumption of unstable $O^-$ ions is based on the experience with $La_{2-x}Ae_xCuO_4$, where  doped holes reside pairwise in $O$ atoms, instead of individually in $O^-$ ions.\cite{1} Coulomb repulsion spreads the double holes in under-annealed $YBa_2Cu_3O_{6+y}$, harbored by $O$ atoms, to form an $O$ superlattice. The resulting charge-order stripes then must have a square-root dependence on oxygen doping akin to Eq. (1), to be resumed shortly.

\section{ANTIFERROMAGNETISM}

The parent crystals $La_2CuO_4$ and $YBa_2Cu_3O_6$ are Mott insulators with antiferromagnetic order of the copper ions' magnetic moments, $\mathbf{m}(Cu^{2+})$, in the $CuO_2$ planes.  
Due to their closed electron shells, neither the oxygen ions contribute to magnetism, $\mathbf{m}(O^{2-})=0$, nor the copper ions in the $Cu(1)$ planes, $\mathbf{m}(Cu^+)=0$.
The stability of the parent crystals' 3D-AFM is dominated by the magnetic interaction between the $CuO_2$ planes. 
Because of orthogonal $\mathbf{m}(Cu^{2+})$ orientation in the
 widely separated, neighboring $CuO_2$ planes, compounded by the staggering of unit cells, 3D-AFM is relatively weak in $La_{2}CuO_{4}$. It collapses when $Ae$-doped by $x_{N0} = 0.02$---a very small N\'{e}el concentration. [This is the quantity that affects the offset value in Eq. (1), $\check{p} \le x_{N0}$.] More $Ae$-doping, $x > x_{N0}$, leads to stripe formation in $La_{2-x}Ae_xCuO_4$, with incommensurability $q_{c,m}(x)$, as given by Eqs. (1-3).

In $YBa_2Cu_3O_6$, by contrast, the $CuO_2$ planes that sandwich the central $Y^{3+}$ planes are close and aligned, which gives rise to strong 3D-AFM. [The separation of neighboring $\mathbf{m}(Cu^{2+})$ moments along the $c$-direction is actually less than in the $CuO_2$ plane.]
Oxygen doping by embedding process \#1 generates additional $\mathbf{m}(Cu^{2+})$ moments in $CuO$ chains along the $b$ direction.
They upset the 3D-AFM of the host and cause its collapse at $y_{N0} \simeq 0.30$ (but $0.25 < y_{N0} < 0.30$ when not well-annealed).
With more oxygen doping, $y_{N0}< y < 0.5$, magnetization stripes are observed {\it only} if the crystal is insufficiently annealed, as mentioned. Their incommensurability rises from zero in a square-root dependence on doping (see Fig. 1, bottom).
As noted, a square-root rise of the incommensurability is a signature of a superlattice of  \emph{defect charges}. 
In the present case the superlattice is formed by lattice-site $O$ atoms in the $CuO_2$ planes, generated by process \#2, carrying double holes. Their magnetic moments, $\mathbf{m}(O)$, aligned in antiferromagnetic order, give rise to magnetization stripes, observed to be parallel to the crystal's $a$-axis. 
 Taking into account that the unit cell of $YBa_2Cu_3O_{6+x}$ has \emph{two} $CuO_2$ planes, the incommensurability of magnetization stripes depends on oxygen doping, akin to Eq. (1),  as
\begin{equation}
q_m(y)  = \frac{\sqrt{2}}{4}\sqrt{\frac{g}{2} (y - \check{y})}\;, \;\;\;\;\; 0.25 < y < 0.50  \; .
\end{equation}
Here $g$ is the fraction of doped oxygen atoms that are embedded by process \#2. From the three known data points of magnetization stripes,\cite{2,3,4,5} the values of the offset $\check{y}$ and the fraction $g$ can be determined (see Appendix B). The result is $\check{y} = 0.27$, close to the collapse of 3D-AFM at $y_{N0} \approx 0.25$, and $g=0.18$. The latter quantity means that between one-in-five and one-in-six doped oxygen atoms are embedded by the hole-generating process \#2. 
Where are the charge-order stripes of double incommensurability
that accompany the magnetization stripes? By analogy with doped lanthanum cuprates, $La_{x}Ae_xCuO_4$, one can infer that they are, in this doping range, below the present detection limit.\cite{1}
The crystals where magnetization stripes have been observed, $0.30 \le y \le 0.45$, were annealed in the span of a day.\cite{2,3,4,5} However, such stripes are absent in crystals annealed over three months (and where $y_{N0} = 0.31$).\cite{6} 
This implies that the filling of $CuO$ chains by ``electron borrowing'' from the $CuO_2$ planes (process \#2) is a transitory phenomenon. But if the sample is cooled too fast, the attending magnetization stripes remain ``frozen-in.'' 

\section{ORTHO-STRUCTURE OF OXYGEN CHAINS}
Figure 1 shows that the magnetization stripes from incomplete annealing terminate when $y \rightarrow 0.5$. Oxygen doping of $YBa_2Cu_3O_{6+y}$ by embedding processes \#1 or \#2 is exhausted at $y=0.5$. 
Larger oxygen doping, $y>0.5$, affects both the $Cu(1)$ plane and each of the (two) adjacent $CuO_2$ planes, 
\begin{equation}
y = \chi^{O(1)}(y) + 2\delta^{O(2)}(y)\;.
\end{equation}
The two contributions are determined by two processes, labelled \#3 and \#4, concerning oxygen embedding into $CuO$ chains, $O(1)$, and respectively, into $CuO_2$ planes, $O(2)$.
\medskip  

\noindent \textbf{\#3.} In the $Cu(1)$ plane, enriching oxygen {\it atoms}, $O$---instead of $O^{2-}$ ions---now fill more than every other $CuO$ chain. The chains---always along the $b$-direction---are called ``ortho-$N$ oxygen ordered'' when they have a periodicity $Na$ in the  $a$-direction.\cite{29} 
The observed chain orderings are listed in Table I. The simplest cases are ortho-II with f-e sequence (full and empty chains alternating along the $a$-direction) and ortho-III with f-f-e sequence. The other orderings can be regarded as combinations thereof: V = II + III and VIII = III + II + III. 

Although the doping fraction $y$ can be accurately established by the proportion of ingredients to crystal growth, the \textit{partition} of doped oxygen to the $CuO$ chains and to the $CuO_2$ planes, Eq. (8), is not well known. Under the the idealized assumption that the chains are completely full or empty---which is best approximated at prominent doping levels $\underline{y}$---along with the ``0.5-watershed,''
we can use the perfect fill ratio, $\phi_{\infty}^{O(1)}$=N$_f$/N, where N$_f$ is the number of full chains in an ortho-N oxygen order. With $\chi^{O(1)}(\underline{y}) \approx \phi_{\infty}^{O(1)}(\underline{y})$, Eq. (8) gives
\begin{equation}
\delta_{0.5}^\infty(\underline{y}) \approx \frac{ \underline{y} -  \phi_{\infty}^{O(1)}(\underline{y})}{2}\;,
\end{equation}
listed in Table I. At $y=0.5$, the $Cu(1)$ plane is almost perfectly covered by ortho-II $CuO$ chains, as attested by very high correlation lengths, $\xi_b(\underline{y})/b_0=112$ and $\xi_a(\underline{y})/a_0=39$.\cite{30} 
Other cases of $CuO$ chains, while long, are not completely filled, as indicated by the observed chain-order correlation length $\xi_b(\underline{y})/b_0$ in Table I. Nevertheless, the $\delta_{0.5}^\infty(\underline{y})$ values in the table give a rough estimate and overview of the density of oxygen atoms in the $CuO_2$ planes.
 
\begin{table}[ht!]
\begin{tabular}{|p{2.65cm}|p{1.2cm}|p{1.6cm}|p{2.05cm}|p{2.4cm}|p{2.4cm}|p{1.5cm}|p{1.5cm}|}
 \hline  \hline
oxygen $\;\;\;\;\;\;\;\;$ content &promi- nent $\underline{y}$&chain $\;\;\;\;\;\;\;$ ortho-N&chain $\;\;\;\;\;\;\;$ sequence& $\phi_{\infty}^{O(1)}(\underline{y})$ in the Cu(1) plane& 
$\delta_{0.5}^\infty(\underline{y})$ in each $\;\;\;\;\; CuO_2$ plane&
$\xi_b(x)/b_0$ chains&$\xi_{\tau}(y)/\tau_0$ stripes\\
 \hline  \hline
$0.35 \le y \le 0.50$& &II =&f-e (spotty)& & &$\;\;\;\;$44&$\;\;\;\;$10 \\
$0.50 < y \le 0.6$&$\;\;$1/2&II =&f-e&$\;1/2\;=\;0.50$&$\;\;\;\;$ $\;\;\;0.00$&$\;\;\;\;$61&$\;\;\;\;$17 \\
$y=0.62$&  &(II\&) V=&f-e-f-f-e&$\;3/5\;=\;0.60$&$\;\;\;\;\;\;\;\;0.01$&$\;\;$& \\
$y\simeq 0.67$&$\;\;$2/3&VIII =&f-f-e-f-e-f-f-e
&$\;5/8\;=\;0.625$&$\;\;\;\;\;\;\;\;0.021$&$\;\;\;\;36$&$\;\;\;\;$26 \\
$0.72 \le y \le 0.82$&$\;\;$3/4&III =&f-f-e&$\;2/3\;=\;0.667$&$\;\;\;\;\;\;\;\;0.042$&$\;\;\;\;30$&$\;\;\;\;$11\\
$0.92$&&IV =&f-f-f-e&$\;3/4\;=\;0.75$&$\;\;\;\;\;\;\;\;0.085$&&\\
 \hline   \hline
\end{tabular}
\caption{$CuO$ chains in the $Cu(1)$ planes of $YBa_2Cu_3O_{6+y}$ with ortho-N ordering along the $a$-direction of full (f) and empty (e) chains along the $b$-direction at prominent doping levels $\underline{y}$. The quantity $\phi_{\infty}^{O(1)}$=N$_f$/N is perfect fill ratio and
$\delta_{0.5}^\infty(\underline{y})$, Eq. (9), is the density of $O$ atoms embedded in each $CuO_2$ plane at prominent doping levels $\underline{y}$, based on idealized assumptions. 
The observed correlation lengths (Refs. 8-11, 14, 15) are denoted by $\xi$, with $b$-axis lattice constant $b_0$ and $\tau = a,b$.}

\label{table:1}
\end{table}

\section{CONVENTIONAL VIEW OF HOLE DOPING BY OXYGEN ENRICHMENT}
Charge-order stripes are observed in the $CuO_2$ planes of $YBa_2Cu_3O_{6+y}$ for oxygen doping $y>0.5$. 
(Transcending the assumption of the ``0.5-watershed,'' weak stripes appear already in the $0.45 < y < 0.5$ interval). The part of oxygen doping that is not consumed by $CuO$ chain filling, but causes charge-order stripes in each $CuO_2$ plane, is denoted as $\delta(y) \equiv \delta^{O(2)}(y)$, Eq. (8).
\textit{Common wisdom} has it, that each stripe-forming oxygen atom, $O_{\delta}$, upon taking two electrons from an $O^{2-}$ ion in an $CuO_2$ plane, resides as an interstitial ion, $O_{\delta}^{2-}$, in the so-called charge-reservoir layer (which comprises all planes of the unit cell other than $CuO_2$). This amounts to hole doping of the $CuO_2$ plane:
\newline .
\newline \noindent .$\;\;\;\;\;\;\;\;\;\;\;\;\;\;\;| .......\;O^{2-}Cu^{2+}O^{2-}................|\; \rightarrow  \;|.......\;O^{2-}Cu^{2+}O^{2-}+2e^+.............|$
\newline
\noindent $O_{\delta\;}\;\; +\;\;\;\;|$......Charge-reservoir layer......$|\;\rightarrow  \;|$...Charge-reservoir + $O_{\delta}^{2-}$ layer...$|$
\bigskip \medskip

\noindent If, like in $La_{2-x}Ae_xCuO_4$, the doped holes react as $2e^+ + O^{2-} \rightarrow O$ to form lattice-site $O$ atoms,
their superlattice would give rise to charge-order and magnetization stripes with a square-root dependence of $q_{c,m}(\delta)$, akin to Eqs. (1, 3). However, \textit{no such stripes are observed} in $YBa_2Cu_3O_{6+y}$ (for $y>0.5$), but $q_c(\delta) \approx 0.3$ and \textit{no} magnetization stripes at all (see Fig. 1). Therefore, the conventional view cannot be correct. 

\section{OXYGEN EMBEDDING IN THE $\mathbf{CuO_2}$ PLANES}
Concerning the implementation of stripe-forming oxygen for $y>0.5$, five observations shall be pointed out, all visible in Fig. 1: (i) the onset of charge-order stripes with a \emph{non-zero} value, $q_c^{ons} \ne 0$; (ii) the stripes' incommensurability $q_c(y)$ \emph{not} having a rising square-root dependence; (iii) the near \emph{constancy} of $q_c(y)$, \textit{decreasing} slightly at larger $y$; (iv) the specific range of $q_c(y) \simeq 0.3$ r.l.u., and (v) the \emph{lack} of accompanying magnetization stripes. 
The five observations give rise to five implications: 
(i) The non-zero onset suggests a \emph{different} mechanism of stripe formation than of the magnetization stripes at $y\approx0.25$ with $q_m^{ons} = 0$ (and as known from $q_{c,m}^{ons} = 0$ of the `214' compounds).
(ii) The absence of a rising square-root dependence of $q_c(y)$ indicates that these stripes are \emph{not} formed by entities with lattice-defect charges, like double holes in lattice-site $O$ atoms, that would spread by Coulomb repulsion. 
(iii) Leaving the slight $y$-dependence of $q_c(y)$ aside (until Sect. IX), a {\it constant} incommensurability is consistent with the aggregation of \emph{neutral} stripe-forming entities.
(iv) Averaging over the small difference of lattice constants, $s_0 = (a_0 + b_0)/2$, and incommensurabilities, $\overline{q}_c(y) = [q_c^a(y) + q_c^b(y)]/2$, the stripe-forming entities aggregate with a spacing $\overline{L} = s_0/\overline{q}_c$. This provides a clue as to their length. Its value is roughly $\overline{L} \approx s_0/0.3 \approx 3 s_0 $, that is, about three lateral lattice constants.
(v) The absence of accompanying magnetization stripes indicates that the stripe-forming entities have no uncompensated magnetic moment, $\mathbf{m} = 0$. With these considerations in mind, the following process of oxygen implementation 
is proposed:

\medskip
\noindent \textbf{\#4.}
The enriching oxygen is implemented by embedding (neutral) $O$ atoms
in the empty centers of $O^{2-}$ quartets in the $CuO_2$ planes, as illustrated at the center of Fig. 4. We want to call these sites ``\textit{pores}'' and symbolize an atom's presence in a pore with an over-ring, \textit{viz.}, $\mathring{O}$. Benefiting from the attending gain in binding energy, the interstitial oxygen atom combines with two $O^{2-}$ neighbors along the $a$- or $b$-axis to an \emph{ozone} molecule ion, 
\begin{equation}
 O^{2-}+ \; \mathring{O}\; + \;  O^{2-}  \rightarrow \;
  \ddot{O}\mathring{O}\ddot{O}  \;, \;\;\;\;\;  y > 0.5 \; ,
\end{equation}
illustrated in the second row  of Fig. 4 (from top and bottom). The double over-dots in Eq. (10) signify ionic charges, $\ddot{O} \equiv O^{2-}$. Instead of being isolated, it is energetically favorable that the ozone ions,  $\ddot{O}\mathring{O}\ddot{O}$, align in the same row between intermediate oxygen ions of the host, $O^{2-} =\mathbf{\ddot{O}}$ (marked bold), to form sequences
$...\mathbf{\ddot{O}}\ddot{O}\mathring{O}\ddot{O}\mathbf{\ddot{O}}\ddot{O}\mathring{O}\ddot{O}\mathbf{\ddot{O}}\ddot{O}\mathring{O}\ddot{O}\mathbf{\ddot{O}}\ddot{O}\mathring{O}\ddot{O}\mathbf{\ddot{O}}...$, with distances $\overline{L}$ (center-to-center) apart, shown by the red double arrow in Fig. 4, second row from the bottom.
Viewed differently, the quantity $\overline{L}$ also signifies the length that is occupied by an $\mathbf{\ddot{O}}\ddot{O}\mathring{O}\ddot{O}\mathbf{\ddot{O}}$ motif, as can be seen in the second row from the top of Fig. 4.
By forming the central ozone ion, the {\it five} nuclei of $\mathbf{\ddot{O}}\ddot{O}\mathring{O}\ddot{O}\mathbf{\ddot{O}}$ extend slightly more than the {\it four} nuclei of an $\mathbf{\ddot{O}}\mathbf{\ddot{O}}\mathbf{\ddot{O}}\mathbf{\ddot{O}}$
sequence of the host crystal, as in row 4 (compare red and black double arrows).
Figure 4 illustrates the case of oxygen doping $y \simeq 0.55$ where $\overline{q}_c = 0.32$
and $\overline{L}= s_0/0.32 = 3.125  s_0 = \frac{25}{8}s_0 = \frac{3 \times 8 +1}{8} s_0 $. 
Thus a train of eight contiguous $\ddot{O}\mathring{O}\ddot{O}\mathbf{\ddot{O}}$ motifs ($8 \times 3 = 24$ nuclei) would extend over the length of $8 \times 3.125 = 25$ ideal lattice sites. A reason for the out-of-plane tilt of the central ozone ion will be given shortly.

The magnetic moment of the ozone ion, $\mathbf{m}(\ddot{O}\mathring{O}\ddot{O}) = 0$, is determined by the spin configuration  of its constituting atomic $2p$ subshells, $[\uparrow]$ $[\uparrow\downarrow]$ $[\uparrow\downarrow]$--$[\uparrow\downarrow]$ $[\uparrow\downarrow]$ $[\uparrow\downarrow]$--$[\uparrow\downarrow]$ $[\uparrow\downarrow]$ $[\downarrow]$.
This is in contrast to the spin configuration 
$[\uparrow\downarrow]$ $[\uparrow]$ $[\uparrow]$  of a \emph{lattice}-site $O$ atom, generated by  ``electron borrowing'' in the embedding process \#2 (at oxygen doping $y < 0.5$), with $\mathbf{m}(O) \ne 0$. 
The latter is caused by Hund's rule of maximal multiplicity which is commonly interpreted

\includegraphics[width=6.25in]{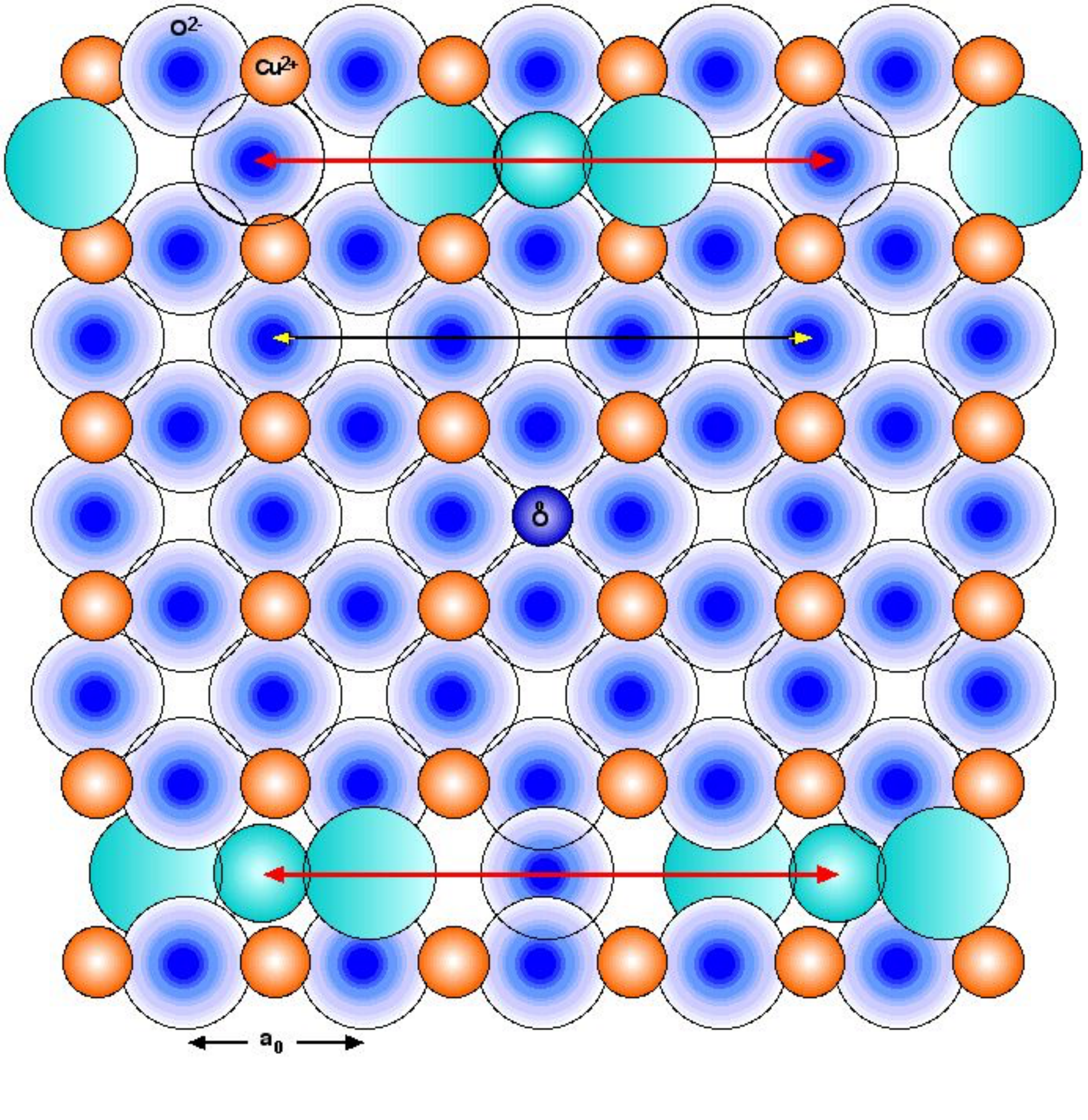}
\footnotesize 
\noindent FIG. 4. Doped oxygen in the $CuO_2$ plane of $YBa_2Cu_3O_{6+y}$ ($y \simeq 0.55$), shown schematically for an oxygen atom at the stages of embedding (center, sixth row) and final relaxation (second row from top and bottom). In the interstitial space (``pore'') between four $O^{2-}$ ions (denoted $\ddot{O}$ for short), the embedded oxygen atom (denoted $\mathring{O}$) bonds with two $O^{2-}$ neighbors, here in the $a$-direction, to form an ozone molecule ion, $\ddot{O}\mathring{O}\ddot{O}$ (turquoise color).
Linked by an intermediate oxygen ion each (marked bold, $\mathbf{\ddot{O}}$), the ozone molecules line up to form trains of $\ddot{O}\mathring{O}\ddot{O}\mathbf{\ddot{O}}$ motifs. The length of a motif is $\overline{L} \simeq 3.125\; a_0$ (red double arrow), slightly larger than $3a_0$ (black double arrow). Its reciprocal gives the incommensurabiliy of charge-order stripes, $\overline{q}_c =1/\overline{L} \simeq 0.32$ r.l.u. 

\normalsize

\pagebreak
\noindent as assistance of parallel spins in reducing Coulomb repulsion between the $O$ atom's lone $p$ orbitals. In the $\ddot{O}\mathring{O}\ddot{O}$ molecule, by contrast, the lone $p$ orbitals, located about the terminal nuclei, are far apart. Thus no assistance from parallel spins is necessary to reduce their much smaller Coulomb repulsion. Instead, it is energetically favorable that the lone $p$ orbitals profit in energy from magnetic attraction of antiparallel spins.

Besides the arguments based on stripe incommensurability---non-zero onset, no square-root dependence, nearly constant $q_c(y) \approx 0.3$, no magnetization stripes---is there other experimental support for the proposed stripe formation in $YBa_2Cu_3O_{6+y}$ ($y>0.5$) by interstitial oxygen, $\mathring{O}$, in the $CuO_2$ planes?
A hard-X-ray investigation of ion displacements in $YBa_2Cu_3O_{6.54}$, evaluated from intensities of \emph{all} available charge-order satellite peaks, showed an accordion-like, incommensurate folding of the $CuO_2$ planes, with distinct out-of-plane displacements.\cite{17}
For example, for stripes along the $a$-direction, the four $O^{2-} = \ddot{O}$ ions at positions $(0,\frac{1}{2},0)$, $(1,\frac{1}{2},-c)$, $(2,\frac{1}{2},+c)$, $(3,\frac{1}{2},0)$---as in row 4 of Fig. 4---
look like

\bigskip 

$\;\;\;\;\;\;\;\;\;\;\;\;\;\;\;\;\;\;\;\;\;\;\;\;\;\;\;\;\;\;\;\;\;\;\;\;\;\;\;\;\;\;\;\;\;\;\;\;\;\;\;\;\;\;\;^c{\uparrow} \;
\mathbf{\ddot{o}}$\textbackslash$_{\ddot{O}}$/$^{\ddot{O}}$\textbackslash
$\mathbf{\ddot{o}}\; \rightarrow _a \;$ 
\bigskip 

\noindent in the $(a,\frac{1}{2},c)$ plane, spanned by the axial arrows.
The bold $\mathbf{\ddot{O}}$ ions are in the $CuO_2$ plane ($c=0$), and the $\mp c$-displacements of the middle $\ddot{O}$ ions are exaggerated. The folding pattern is continued periodically along $a$.

What would compel the host $CuO_2$ planes, upon oxygen doping $y > 0.5$, to fold like this?
As a common experience, a thin piece of material---say a ribbon or a foil---folds when enlarged while being confined to one or two dimensions, by extending in a perpendicular dimension. This basic phenomenon may also be exhibited by the $CuO_2$ planes of $YBa_2Cu_3O_{6+y}$. The corresponding ``enlargement'' may well be caused by interstitial $\mathring{O}$ atoms that bond with neighboring $O^{2-}$ ions to form ozone ions, $\ddot{O}\mathring{O}\ddot{O}$. 
Space constraints along the stripes' direction, here along $a$, make these ions tilt, causing a folding pattern like

\bigskip \medskip
$\;\;\;\;\;\;\;\;\;\;\;\;\;\;\;\;\;\;\;\;\;\;\;\;\;\;\;\;\;\;\;\;\;\;\;\;\;\;\;\;\;\;\;\;\;\;\;\;\;\;\;\;\;\;\;^c{\uparrow} \;
\mathbf{\ddot{o}}$\textbackslash$_{\ddot{O}}$\textit{o}$^{\ddot{O}}$\textbackslash$\mathbf{\ddot{o}}$
$\; \rightarrow _a\;.$

\bigskip \medskip


\noindent The interstitial $\mathring{O}$ atom is at the pore position, $(\frac{3}{2},\frac{1}{2},0)$, like in row 6 of Fig 4. (Graphical constraints prohibit the over-ring to be shown in the last folding pattern.) Qualitatively, the similarity of $\ddot{O}$ ion displacements in both folding patterns is striking. While the reason for folding the host $CuO_2$ plane (without $\mathring{O}$, top pattern) is not obvious, space requirement makes it necessary when $\mathring{O}$ is present (bottom pattern).

Getting back to the top folding pattern of host ion displacements, obtained from the X-ray investigation, not only do the $O^{2-}$ ions in the $(a,\frac{1}{2},c)$ plane (row 4 in Fig. 4) have large $\mp c$ displacements, but also their neighbors in the $CuO_2$ plane (rows 3 and 5). The next diagram shows these neighbors, displayed in the corresponding $(a,\frac{1}{2} \mp \frac{1}{2},c)$ planes that bracket the central $(a,\frac{1}{2},c)$ plane. (The middle line, labelled `$b=\frac{1}{2}$,' is a repetition of the top folding pattern above.) 
The bracketing $(a,\frac{1}{2} \mp \frac{1}{2},c)$ planes contain both
$Cu^{2+} \equiv \ddot{C}$ and $O^{2-} \equiv \ddot{O}$ ions, with separation $\ddot{C}$-$\ddot{O} = a_0/2$ in the $a$-direction (see rows 3 and 5). For the sake of clarity, the $Cu^{2+} = \ddot{C}$ and $O^{2-} = \ddot{O}$ ions are displayed separately (top and bottom line of the following diagram, labelled `$b=0,1$,'), with the understanding that the ions of \emph{both} lines bracket the center line on both sides (as in rows 3 and 5 of Fig. 4). 

\medskip 
$b=0,1\;\;\;\;\;\;\;\;\;\;\;\;\;\;\;\;\;\;\;\;\;\;\;\;\;\;\;\;\;\;\;\;\;\;\;\;\;\;\;\;\;\;\;^c{\uparrow} \;
\mathbf{\ddot{c}}/^{\ddot{C}}$\textbackslash$ _{\ddot{C}}/\mathbf{\ddot{c}}
\;\; \rightarrow _a $ 

$b=\frac{1}{2}\;\;\;\;\;\;\;\;\;\;\;\;\;\;\;\;\;\;\;\;\;\;\;\;\;\;\;\;\;\;\;\;\;\;\;\;\;\;\;\;\;\;\;\;\;\;^c{\uparrow} \;
\mathbf{\ddot{o}}$\textbackslash$_{\ddot{O}}$/$^{\ddot{O}}$\textbackslash$\mathbf{\ddot{o}}$
$\; \rightarrow _a $ 

$b=0,1\;\;\;\;\;\;\;\;\;\;\;\;\;\;\;\;\;\;\;\;\;\;\;\;\;\;\;\;\;\;\;\;\;\;\;\;\;\;\;\;\;\;\;^c{\uparrow} \;\;\;
^{\ddot{O}}$\textbackslash$\mathbf{\ddot{o}}$\textbackslash$_{\ddot{O}}$
$\;\;\; \rightarrow _a $ 

\medskip 
\noindent Now we can also see out-of-plane $\pm$ displacements of the middle $\ddot{C}$ and terminal $\ddot{O}$ ions in the $b=0,1$ lines: They are opposite to the $\mp c$ displacements of the middle $\ddot{O}$ ions in the $b=\frac{1}{2}$ line---with a ``butterfly pattern of oxygen shear displacements''\cite{17}---as if they want to avoid each other. Why?

The pattern of host ion displacement makes sense when we insert an interstitial $\mathring{O}$ atom in the central line, $b=\frac{1}{2}$, shown in the next folding pattern. Space constraints along the $a$-direction causes the ensuing ozone ion, $\ddot{O}\mathring{O}\ddot{O}$, to tilt out of the $CuO_2$ plane. The neighbor ions in the $b=0,1$ lines accommodate the tilt by getting out of the way.

\medskip 
$b=0,1\;\;\;\;\;\;\;\;\;\;\;\;\;\;\;\;\;\;\;\;\;\;\;\;\;\;\;\;\;\;\;\;\;\;\;\;\;\;\;\;\;\;\;^c{\uparrow} \;
\mathbf{\ddot{c}}/^{\ddot{C}}$\textbackslash$ _{\ddot{C}}/\mathbf{\ddot{c}}
\;\; \rightarrow _a $ 

$b=\frac{1}{2}\;\;\;\;\;\;\;\;\;\;\;\;\;\;\;\;\;\;\;\;\;\;\;\;\;\;\;\;\;\;\;\;\;\;\;\;\;\;\;\;\;\;\;\;\;\;^c{\uparrow} \;
\mathbf{\ddot{o}}$\textbackslash$_{\ddot{O}}$\textit{o}$^{\ddot{O}}$\textbackslash$\mathbf{\ddot{o}}$
$\; \rightarrow _a $ 

$b=0,1\;\;\;\;\;\;\;\;\;\;\;\;\;\;\;\;\;\;\;\;\;\;\;\;\;\;\;\;\;\;\;\;\;\;\;\;\;\;\;\;\;\;\;^c{\uparrow} \;\;\;
^{\ddot{O}}$\textbackslash$\mathbf{\ddot{o}}$\textbackslash$_{\ddot{O}}$
$\;\;\; \rightarrow _a $ 

\medskip 

\noindent As a technical note, charge-order stripes in the $b$-direction of in $YBa_2Cu_3O_{6.54}$ are much more prevalent than along $a$ (see Sect. IX). Consequently, the X-ray evaluation of host-ion displacements is much more accurate for $b$-stripes than for $a$-stripes. The present explanation of out-of-plane folding for $a$-stripes was done for ease of display. All conclusion hold equally for $b$-stripes.

The X-ray investigation of host-ion displacements\cite{17} showed no direct evidence of excess oxygen in the $CuO_2$ planes. Subject to symmetries and certain constraints, the evaluation of the diffraction intensities was based on the best fit of all displacements of all host ions in the unit cell. Mirror symmetry with respect to the $Cu(1)$ planes was assumed and group-theoretical constraints were invoked. However, no allowance for interstitial oxygen in the $CuO_2$ planes was provided in the evaluation---thus none \emph{could} be found. Nevertheless, the host-ion displacements from the investigation are, on a qualitative level, compatible with the presence of interstitial oxygen in the $CuO_2$ planes and the observed folding is compatible with out-of-plane tilts of $\ddot{O}\mathring{O}\ddot{O}$ ions. It would be desirable to revaluate the X-ray intensities\cite{17} with allowance for interstitial oxygen in the $CuO_2$ planes.

\section{INCOMMENSURABILITY AND AXIAL PREFERENCE OF STRIPES}

In order to go beyond the simplified assumption of constant incommensurability, treated in the previous section, three observations must be taken into account.
(i) Figure 1 shows that the $q_c$ data decrease by 13\% from $q_c^b(0.45)=0.335 \simeq\frac{8}{24}$ to $q_c^a(0.90)=0.292 \simeq \frac{8}{28}$. 
This means that the $\mathbf{\ddot{O}}\ddot{O}\mathring{O}\ddot{O}\mathbf{\ddot{O}}$ length increases such that a train of eight contiguous $\ddot{O}\mathring{O}\ddot{O}\mathbf{\ddot{O}}$ motifs extends with increasing $y$ over a distance from 24 to 28 lattice sites, respectively.
\newline (ii) Figure 1 also shows that for a given oxygen doping $y$, stripe incommensurability along the $b$-axis is always larger than for stripes along the $a$-axis, $2(q_c^b-q_c^a)/(q_c^b+q_c^a)\approx 3\%$.
\newline (iii) At $y\approx0.55$, the peak intensity of X-rays diffracted by charge-order stripes is dominant for stripes along the $b$-direction, but very weak for stripes along $a$.\cite{14} (The intensity of stripes is not indicated in Fig. 1.) At $y=0.67$, peak intensities are about equal for stripes along $b$ and $a$.\cite{14}
With $y > 0.67$, the peak intensity gets increasingly stronger for $a$-stripes but weaker for $b$-stripes.\cite{14} This means, there is an axial preference of the stripes, depending on oxygen doping.
The three observations raise three questions: (i) Why does 
$\overline{q}_c(y)$ {\it decrease} with increasing doping? 
(ii) Why is always $q_c^b(y) > q_c^a(y)$ ? (iii) What is the reason for the change of axial preference with increasing oxygen doping?

From the incommensurability $q_c^b=0.335$ of the very weak stripe signal, measured at $y=0.45$, we can infer that an \emph{isolated} $\mathbf{\ddot{O}}\ddot{O}\mathring{O}\ddot{O}\mathbf{\ddot{O}}$ motif, with its tilted central ozone ion, projects on the $b$-axis to a length $L^b = b_0/q_c^b=3b_0$. When, with increasing oxygen doping, several $\ddot{O}\mathring{O}\ddot{O}\mathbf{\ddot{O}}$ motifs axially connect, van-der-Waals attraction---specifically between their out-of-plane $\ddot{O}$  members (marked bold in the diagram)---reduce the ozone tilt,

\bigskip

$\;\;\;\;\;\;\;\;\;\;\;\;\;\;\;\;\;\;\;\;\;\;\;\;\;\;\;\;\;\;\;\;\;\;\;\;\;\;\;\;\;\;\;\;\;\;\;\;\;\;\;\;\;\;\;^c{\uparrow} \;
\ddot{o}_{\ddot{O}}$o$^{\mathbf{\ddot{O}}}\ddot{o}_{\mathbf{\ddot{O}}}$o$^{\ddot{O}}\ddot{o}$$\; \rightarrow _{a,b}\;,$

\medskip
\noindent with subsequent increase of their projected axial length, $\overline{L}(y)$. This explains, qualitatively, the slight decrease of incommensurabilty with oxygen enrichment, $\overline{q}_c(y) = s_0/\overline{L}(y)$. 

For a quantitative assessment, we relax the idealized assumption of the ``0.5-watershed'' which served in introducing oxygen-implementation separately for processes \#1,2 and \#3,4. Experimental evidence from the onset of stripes and of superconductivity indicates that to a small degree processes \#4 commences already at $y < 0.5$. Accordingly, process \#1 extends in the same (small) degree to $y > 0.5$.
Motivated by the quest for a common description for \emph{all} ($Y$-, $Hg$-, $Bi$-based) oxygen-enriched cuprates, we express the dependence of stripe incommensurability on the density $\delta$ of excess oxygen in each $CuO_2$ plane empirically as
\begin{equation}
\overline{q}_c(y) = \overline{q}_c(y_c^{ons}) - \gamma \times [\delta(y)-\delta(y_c^{ons})]\;. 
\end{equation}
Here the superscript {\it `ons'} indicates the onset of charge-order ($c$) stripes.
The onset incommensurability, $\overline{q}_c(y_c^{ons})$, is material-specific, typically in the range $0.35 > \overline{q}_c(y_c^{ons}) >0.25$. For $YBa_2Cu_3O_{6+y}$, $y_c^{ons}=0.45$ and $\overline{q}_c(y_c^{ons})=0.33$. Both the coefficient $\gamma \simeq 0.64$/r.l.u. and the onset density $\delta(y_c^{ons}) \simeq 0.035$ (evaluated in Sect. X) hold for the whole \textit{family} of oxygen-enriched cuprates. Values of $\delta$, calculated with Eq. (11) and $\overline{q}_c(y)$ data from experiment, are listed in Table II. They are systematically larger than the values obtained with idealized assumptions (``0.5-watershed,'' filled $CuO$ chains) from Table I, $\delta(y) > \delta_{0.5}^\infty(y)$.

Turning to the second question---why is $q_c^b(y) > q_c^a(y)$?---one third of the 3\% difference of $q_c^{b,a}$, that is 1\%, is a consequence of orthorhombic anisotropy: The same actual length of an


\begin{table}[ht!]
\begin{tabular}{|p{1.15cm}|p{1.5cm}|p{2.8cm}|p{2.5cm}|p{3.2cm}|p{4.5cm}|}
 \hline  \hline
$\;\;\;y$&$\;\overline{q}_c$ expt.&$\;2\delta_{0.5}^\infty$ $\leftarrow$ Table I&$\;2\delta$ $\leftarrow$ Eq. (11)& $\;p$ $\leftarrow$ Univ. Dome& $\;$Comment\\
 \hline  \hline
$0.31 (1)$&&&&$\;\;\;\;\;\;\;\;\;\;\;0.05$&Onset of superconductivity \\  
$\;0.45$&$\;\;0.33$ &&$\;\;\;\;\;\;\;\;0.07$&$\;\;\;\;\;\;\;\;\;\;\;0.08$&Onset of charge stripes  \\
$\;0.67$&$\;\;0.31$&$\;\;\;\;\;\;\;\;0.04$&$\;\;\;\;\;\;\;\;0.13$&$\;\;\;\;\;\;\;\;\;\;\;0.12$& Maximal stripe intensity \\
$\;0.75$&$\;\;0.303$&$\;\;\;\;\;\;\;\;0.08$&$\;\;\;\;\;\;\;\;0.16$&$\;\;\;\;\;\;\;\;\;\;\;0.14^*$&*Ignoring the dip of $T_c(y)$ \\
$\;0.92$&$\;\;0.297$&$\;\;\;\;\;\;\;\;0.17$&$\;\;\;\;\;\;\;\;0.18$&$\;\;\;\;\;\;\;\;\;\;\;0.16$&$T_{c,max} = 93.5$ K \\

 \hline   \hline
\end{tabular}
\caption{The density $\delta(y)$ of stripe-forming oxygen in the $CuO_2$ planes of $YBa_2Cu_3O_{6+y}$, calculated with Eq. (11) and $\overline{q}_c(\delta)$ data from experiment, exceeds  $\delta_{0.5}^\infty(y)$ from Table I, obtained with idealized assumptions. Values of $2\delta_{0.5}^\infty(y)$ and $2\delta$ are given for comparison with the doped-hole density $p$, obtained with the universal-dome method (discussed in Sect. XIII).}

\label{table:2}  \end{table}

\noindent  $\ddot{O}\mathring{O}\ddot{O}\mathbf{\ddot{O}}$ motif (at a given $y$ value) appears 1\% smaller when expressed in terms of the lattice constant $b_0$ than of $a_0$ because of $2(b_0-a_0)/(b_0+a_0)\approx 1\%$. The remaining 2\% difference (at the same $y$ value) is caused by the presence or absence of \textit{squeeze} from full or, respectively,
empty $CuO$ chains in the $Cu(1)$ plane onto the tilted ozone ions in the $CuO_2$ planes. This can be seen in Fig. 5, which holds for oxygen doping $y\simeq0.55$, when the $CuO$ chains along $b$ are ortho-II oxygen ordered (alternately full and empty). Along the $a$-direction, the tilt
\medskip 

\includegraphics[width=5.8in]{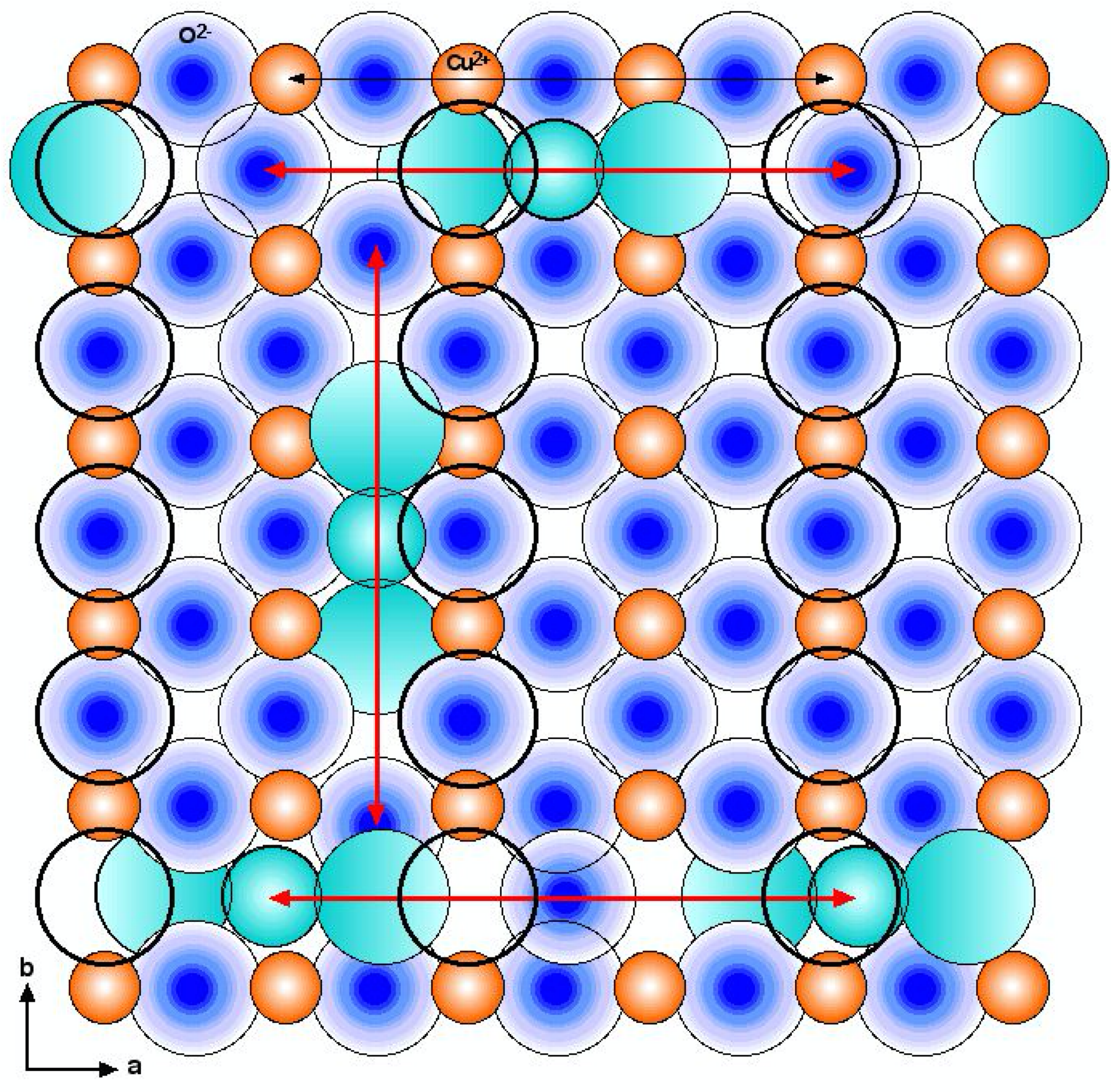}  \footnotesize 

\noindent FIG. 5. Depression of the tilt of ozone ions $\ddot{O}\mathring{O}\ddot{O}$ (turquois) by $CuO$ chains (heavy rings), here in ortho-II order ($y\simeq0.55$): Central squeeze by full $CuO$ chains reduces the tilt of $\ddot{O}\mathring{O}\ddot{O}_a$ ions along $a$. This increases the projected length, $L^a$ (horizontal red arrows), of $\ddot{O}\mathring{O}\ddot{O}\mathbf{\ddot{O}}_a$ motifs. In contrast, the $\ddot{O}\mathring{O}\ddot{O}_b$ ions along $b$ are squeezed by $CuO$ chains only on one side, causing less tilt depression (sideway evasion), such that $L^b < L^a$ and $q_c^b(y)>q_c^a(y)$. At higher doping, less empty $CuO$ chains are available. Rather than being squeezed from full $CuO$ chains on both sides (not shown), $\ddot{O}\mathring{O}\ddot{O}$ ions then prefer to align along $a$ instead of $b$.
\normalsize 

\noindent   of the ozone ions is depressed by the $O^{2-}$ ions in the $CuO$ chains above (or below)---shown by heavy circles---causing a longer projected length $L^a$ of the $\ddot{O}\mathring{O}\ddot{O}\mathbf{\ddot{O}}_a$  motifs.  Figure 5 also shows that, with ortho-II oxygen ordering, tilted ozone ions along the $b$-directions are squeezed by a full $CuO$ chain only on \textit{one side} but not by the empty $CuO$ chain on their other side. Shifting sideways, $b$-oriented $\ddot{O}\mathring{O}\ddot{O}_b$ ions can partly evade the squeeze, resulting 
in $L^b < L^a$. This explains, qualitatively, the finding $q_c^b(y) > q_c^a(y)$.

Turning to the question of axial preference, one-sided depression of tilted ozone ions along 
 \medskip

\includegraphics[width=5.63in]{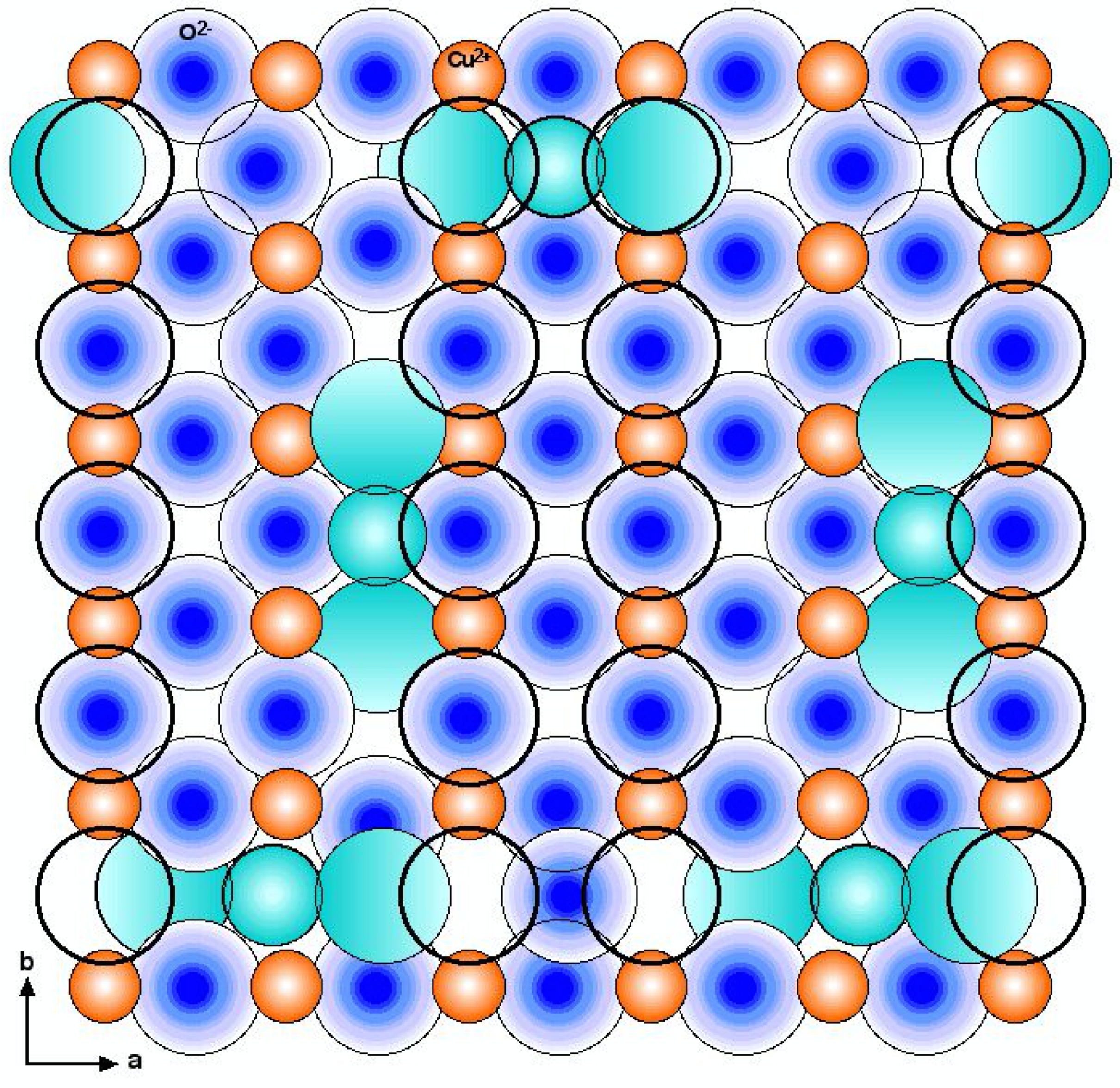}  \footnotesize  
 
\noindent FIG. 6. $CuO_2$ plane of $YBa_2Cu_3O_{6.67}$ with ortho-VIII oxygen order (ffefeffe) and ozone ions. When compressed along $a$, $CuO$ chains (black rings) transmit the increased tilts of $\ddot{O}\mathring{O}\ddot{O}_a$ ions to the next $CuO_2$ plane where some ozone ions rotate, 
$\ddot{O}\mathring{O}\ddot{O}_{a \rightarrow b}$. But when compressed along $b$, no increased tilt of $\ddot{O}\mathring{O}\ddot{O}_b$ ions is translated, as the $CuO$ chains unilaterally slide by. 
This causes the asymmetry of 3D stripe order: Increase of 3D $b$-stripes when compressed along $a$, but no change of 3D $a$-stripes when compressed along $b$.
\normalsize 
\pagebreak

 \noindent $b$ seems to be energetically preferable to central depression of $\ddot{O}\mathring{O}\ddot{O}$ ions along $a$ by full $CuO$ chains. This would explain the prevalence of $b$-stripes in the range $0.45<y<0.6$. What happens at denser ortho-N oxygen structures, for $y>0.6$, when every other $CuO$ chain is no longer empty? Depression of $b$-oriented ozone ions along $b$ on \textit{both} sides by full $CuO$ chains seems to be prohibitive.
 Thus, when less empty $CuO$ chains are available at higher doping, the ozone ions rather align along $a$ instead of $b$. This explains, qualitatively, the shift of the stripes' axial preference from the $b$-direction to the $a$-direction at high oxygen doping.

 How long are the trains of $\ddot{O}\mathring{O}\ddot{O}\mathbf{\ddot{O}}$ motifs? Guided by correlation lengths $\xi_c^{a,b}(y)$ (Tab. I) \newline 
 the trains are $\sim3$ motifs long, $\;\;\;\;\;\;\;\;\;\;\;\;\;\;\mathbf{\ddot{o}}$$_{\ddot{O}}$\textit{o}$^{\ddot{O}}$$\mathbf{\ddot{o}}$$_{\ddot{O}}$\textit{o}$^{\ddot{O}}$$\mathbf{\ddot{o}}$$_{\ddot{O}}$\textit{o}$^{\ddot{O}}$$\mathbf{\ddot{o}}$  $\;\;\;\;\;\;\;\;\;\;\;\;\;\;\;\;\;\;\;\;\;\;\;\;\;\;\;\;\;\;$ at $y=0.5$; 
 \newline rising to a maximum of $\sim6$ motifs, $\;\;\;\;\;\;\;\mathbf{\ddot{o}}$$_{\ddot{O}}$\textit{o}$^{\ddot{O}}$$\mathbf{\ddot{o}}$$_{\ddot{O}}$\textit{o}$^{\ddot{O}}$$\mathbf{\ddot{o}}$$_{\ddot{O}}$\textit{o}$^{\ddot{O}}$$\mathbf{\ddot{o}}$$_{\ddot{O}}$\textit{o}$^{\ddot{O}}$$\mathbf{\ddot{o}}$$_{\ddot{O}}$\textit{o}$^{\ddot{O}}$$\mathbf{\ddot{o}}$$_{\ddot{O}}$\textit{o}$^{\ddot{O}}$$\mathbf{\ddot{o}}$  $\;\;\;\;$ at $y=0.67$; 
 \newline and steadily decrease to $\sim 2$ motifs, $\;\;\;\;\;\;\;\mathbf{\ddot{o}}$$_{\ddot{O}}$\textit{o}$^{\ddot{O}}$$\mathbf{\ddot{o}}$$_{\ddot{O}}$\textit{o}$^{\ddot{O}}$$\mathbf{\ddot{o}}$ 
 $\;\;\;\;\;\;\;\;\;\;\;\;\;\;\;\;\;\;\;\;\;\;\;\;\;\;\;\;\;\;\;\;\;\;\;\;\;$ at $y=0.9$.
 
 Very recently, charge-order stripes have been investigated with RIXS for $YBa_2Cu_3O_{6.67}$ (where the $\ddot{O}\mathring{O}\ddot{O}\mathbf{\ddot{O}}$ trains are longest and about equally oriented along $a$ and $b$), subject to uniaxial compression.\cite{31}
It was found that (i) compression in the $a$-direction increases the intensity, and decreases the peak width, of the RIXS signal from 2D stripes (that is, in the $CuO_2$ planes) along the $b$-direction, and vice-versa. (ii) Compression in the $a$-direction also increases 3D stripe order along $b$, but compression along $b$ leaves 3D stripe order along $a$ unchanged. 

The first finding is readily explained with the present model: Ozone ions avoid constraint in the compressed direction by populating the orthogonal direction. The asymmetry of the second finding is a consequence of the directional asymmetry of the $CuO$ chains: Extending along $b$, the $CuO$ chains transmit increased $\ddot{O}\mathring{O}\ddot{O}$ tilts from one $CuO_2$ plane to the next \textit{differently} for compression along $a$ or $b$, as illustrated in Fig. 6.

\section{STRIPES IN $\mathbf{HgBa_2CuO_{4+\delta}}$}

Mercury-based $HgBa_2CuO_4$ has a tetragonal lattice.\cite{32,33} 
Its unit cell has one $CuO_2$ plane at the center, sandwiched by $BaO$ layers, and terminates with planes of $Hg^{2+}$ ions at the top and bottom (see Fig. 7). When the compound is enriched with oxygen, $HgBa_2CuO_{4+\delta}$, it becomes superconducting, with $T_{c, max} = 98$ K.\cite{32} 
Where will the excess oxygen atoms be embedded? The conventional assignment, based on neutron scattering\cite{32,33} and X-ray diffraction,\cite{34} is at the center of the $Hg$ planes, $(\frac{1}{2},\frac{1}{2},\frac{1}{2}\pm\frac{1}{2})$, called the $O(3)$ position. Evidence 

\pagebreak \includegraphics[width=4.55in]{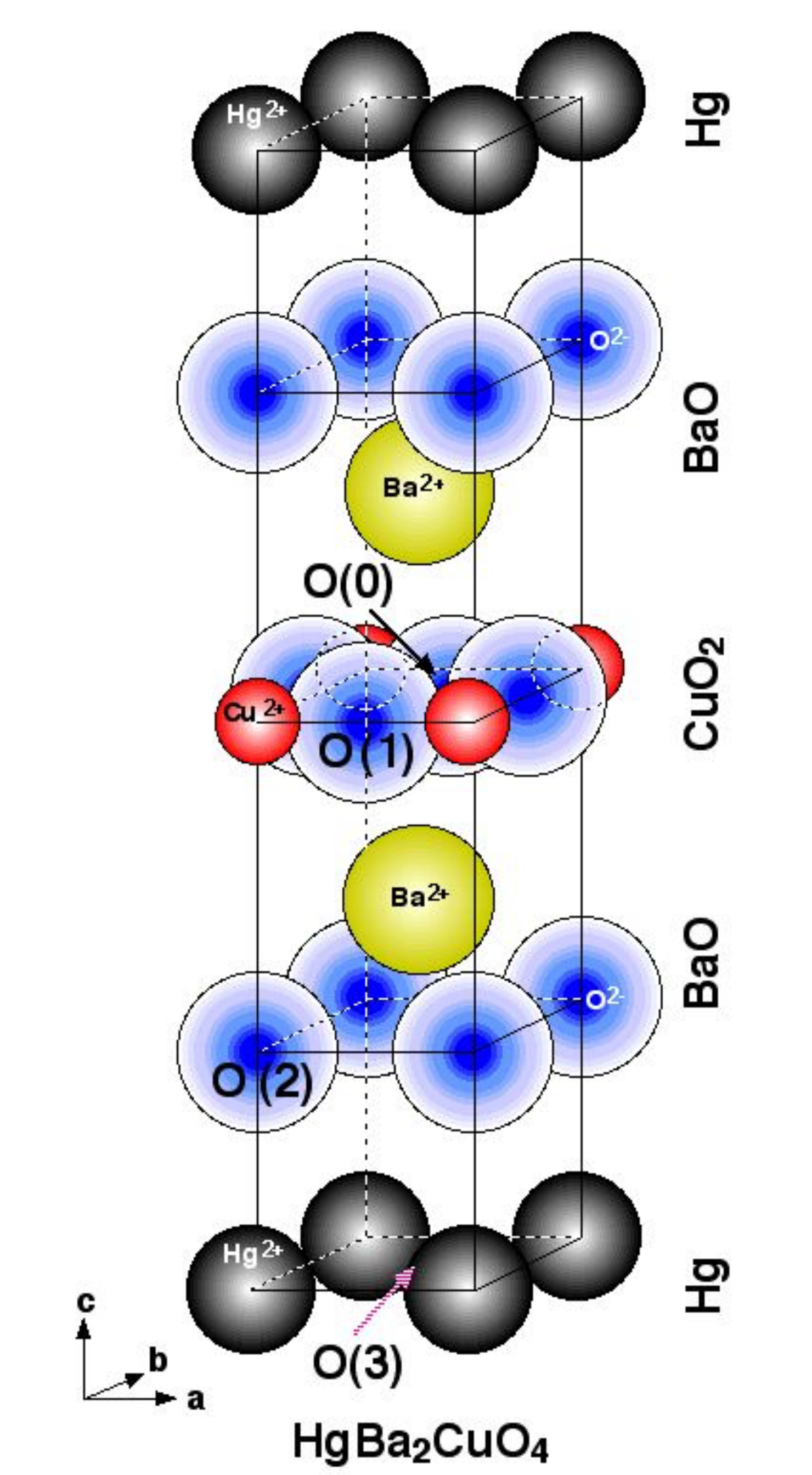} \footnotesize 

\noindent FIG. 7.  Unit cell of $HgBa_2CuO_4$ in hard-sphere display, vertially exploded.  Ion planes and layers are noted on the side. Crystallographic notation of lattice site oxygen,  $O(1)$ and $O(2)$, is given, and of hypothetical sites for interstitial oxygen, $O(3)$ and $O(0)$.
\pagebreak \normalsize 

\noindent   for the $O(3)$ position is indirect: (i) It is supported by Rietveld refinement of the scattering data.\cite{32,33,34} (ii) The decreasing separation of the $Ba$ and $O(2)$  planes with oxygen enrichment is interpreted as the result of charge-transfer from $O(3)$ sites to the $CuO_2$ planes.\cite{32}

Deviating from the conventional assignment, it is assumed here that excess $O$ atoms are embedded in the empty center of $CuO_4$ plaquettes---the ``pores''---as in $YBa_2Cu_3O_{6+y}$, here at the $O(0) \equiv (\frac{1}{2}, \frac{1}{2}, \frac{1}{2})$ position. The deviation is motivated by the quest for a common embedding mechanism of excess oxygen $O_\delta$ in the $CuO_2$ planes of \emph{all} ($Y$-, $Hg$-, $Bi$-based) oxygen-enriched high-$T_c$ cuprates. Supporting arguments for the deviation are: 
 (i) The $O(0)$ position for $O_\delta$ was never probed in the Rietveld refinement of scattering data.\cite{32,33,34} 
 (ii) Although there is ample space for an $O$ atom at the $O(3)$ position between the four $Hg^{2+}$ ions, there is little incentive for it in terms of chemical bonding. However, a gain of cohesive energy would occur at $O(0)$ through formation of an ozone ion, $\ddot{O}\mathring{O}\ddot{O}$, from the excess $O_{\delta}$ atom with two axial $O^{2-}$ neighbors. (iii) Neutron scattering experiments show that the 
 $Ba$-$O(1)$ and $Hg$-$O(2)$ bonds are essentially rigid, but that the $Cu$-$O(2)$ and $Ba$-$O(3)$ distances change considerably with pressure.\cite{33} Due to the rigid $Ba$-$O(1)$ and $Hg$-$O(2)$ layers, compression of the crystal then results in a lesser separation of the $Ba$ and $O(2)$ planes. Accordingly, the observed approach of the $Ba$ and $O(2)$ planes upon enrichment\cite{32} can be interpreted to result from internal pressure exerted by excess $O_\delta$, located at \emph{either} the $O(3)$ \emph{or} $O(0)$ position. If $O_\delta$ resides at $O(3)$, then one would expect that the $Hg$-$O(3)$ distance becomes rigid at higher $\delta$. It is observed, however, that the pressure dependence of the $Hg$-$O(3)$ distance is independent of oxygen enrichment.\cite{33} Combined with the  compressibility of all bonds, the finding is in favor of $O_\delta$ residence at $O(0)$ instead of $O(3)$ sites.

Charge-order stripes appear in  $HgBa_2CuO_{4+\delta}$ at oxygen enrichment $\delta^{ons}_c = 0.035$ with incommensurabilty $q_c^{ons} = 0.292$ when the transition temperature to superconductivity is $T_c(\delta^{ons}) = 55$ K, and disappear at $\delta^{fin}_c=0.085$ with $q^{fin}_c = 0.260$ when $T_c(\delta^{fin}) = 88$ K (Refs. 33, 35). 
Assuming that all enriching oxygen is embedded in the $CuO_2$ planes permits a determination of the familiy-specific rate for stripes in oxygen-enriched $CuO_2$ planes,
\begin{equation}
 \gamma = \frac{q^{fin}_c - q^{ons}_c}{\delta^{fin}-\delta^{ons}} = \frac{0.292 - 0.260}{0.085-0.035} = 0.64\;. 
\end{equation}
Inserting the value into Eq. (11) gives a formula for the dependence of stripe incommensurability on oxygen enrichment of $HgBa_2CuO_{4+\delta}$,
\begin{equation}
q_c(\delta|HgBa_2CuO_{4+\delta}) = 0.292 - 0.64 \;(\delta - 0.035) \;.
\end{equation}
From the observed correlation length of the stripes,\cite{35} $\xi \approx 7.5 a_0$, one can infer that the stripes consist of trains of two motifs, 
$\mathbf{\ddot{o}}$$_{\ddot{O}}$\textit{o}$^{\ddot{O}}$$\mathbf{\ddot{o}}$$_{\ddot{O}}$\textit{o}$^{\ddot{O}}$$\mathbf{\ddot{o}}$ ,
comparable to $YBa_2Cu_3O_{6+y}$ near $y=0.9$.

Beyond enrichment $\delta=0.085$, where $q_c=0.26$, no further stripes are observed in $HgBa_2CuO_{4+\delta}$.
An extrapolation in Ref. 35 of X-ray data to $T_c = 95$ K, as well as Eq. (13) for $\delta=0.10$, gives a \emph{commensurable} value, $q_c = 0.25$ (Bragg peak).
The value $q_c = 0.23$ near optimal doping ($T_c = 95$ K) from another X-ray investigation\cite{36} may originate with a spurious secondary phase.\cite{35}
Streaks connecting the Bragg peaks for higher oxygen enrichment have been attributed to interstitial $O_\delta$ stripes in the $Hg$-$O(2)$ layers.\cite{36,37} 
However, axial stripe formation of ozone molecule ions, consisting of interstitial  $O_\delta$ atoms between lattice-site $O(2)$ ions, is also conceivable.

\section{STRIPES IN BISMUTH-BASED CUPRATES}
Charge-order stripes have been observed in oxygen-enriched, bismuth-based cuprates with single or double $CuO_2$ plane per unit cell.\cite{38,39,40} 
Their incommensurability, $q_c \approx 0.3$, in the bulk of 
$CuO_2$ double-plane $Bi_2Sr_2CaCu_2O_{8+\delta}$---decreasing with more oxygen enrichment (see Table III)---is comparable to $q_c$ of double-plane $YBa_2Cu_3O_{6+y}$. In $Bi_2Sr_{1-x}La_xCuO_{6+\delta}$, of single $CuO_2$ plane per unit cell, the incommensurability, $q_c \approx 0.25$, is also  decreasing with super-oxygenation, and is somewhat less (by $\sim 0.02$) than $q_c$ of single-plane $HgBa_2CuO_{4+\delta}$. From the absence of a square-root dependence of $q_c(\delta)$ on oxygen enrichment and from the absence of magnetization stripes, we can again infer that the stripes are formed by tilted 
\bigskip

\begin{table}[ht!]

\begin{tabular}{|p{5cm}|p{1.2cm}|p{1.2cm}|p{1.5cm}|p{1.4cm}|p{1cm}|}
 \hline  \hline
Compound&$\;T_c$ (K) &$\;\;\;\;p$&$q_c$ (r.l.u)&$\;\;\xi$ (\AA)&$\;$Ref.\\
 \hline  \hline
$Bi_2Sr_{1-x}La_xCuO_{6+\delta}$&$\;\;\;\;15$&$\;0.115$&$\;\;\;0.265$&$\;\;\;\;\;26$&$\;$ 38 \\
$Bi_2Sr_{1-x}La_xCuO_{6+\delta}$&$\;\;\;\;22$&$\;0.130$&$\;\;\;0.257$&$\;\;\;\;\;23$&$\;$ 38 \\
$Bi_2Sr_{1-x}La_xCuO_{6+\delta}$&$\;\;\;\;30$&$\;0.145$&$\;\;\;\;0.243$&$\;\;\;\;\; 21$&$\;$ 38 \\
\hline
$Bi_2Sr_2CaCu_2O_{8+\delta}$&$\;\;\;\;45$&$\;0.09$&$\;\;\;\;0.30$&$\;\;\;\;\; 24$&$\;$ 39 \\
$Bi_{1.5}Pb_{0.6}Sr_{1.54}CaCu_2O_{8+\delta}$&$\;\;\;\;98$&$\;0.16$&$\;\;\;\;0.28$& $\;\;<24$&$\;$ 40 \\
 \hline   \hline
\end{tabular}
\caption{Incommensurability $q_c$ of charge-order stripes in $O$-enriched, $Bi$-based cuprates with single or double $CuO_2$ plane.
The hole density $p$ is obtained from the transition temperature $T_c$ with the universal-dome method (Sect. XIII). The stripes' correlation length is denoted as $\xi$.}
\label{table:3}
\end{table}

\noindent ozone ions. The correlation lengths $\xi$ from Table III indicate trains of about two tilted  
$\ddot{O}\mathring{O}\ddot{O}\mathbf{O}$ motifs, \textit{viz.}, $\mathbf{\ddot{o}}$$_{\ddot{O}}$\textit{o}$^{\ddot{O}}$$\mathbf{\ddot{o}}$$_{\ddot{O}}$\textit{o}$^{\ddot{O}}$$\mathbf{\ddot{o}}$.

$Bi$-based cuprates with \textit{double} $CuO_2$ plane per unit cell show typical features resulting from oxygen enrichment---as encountered in $YBa_2Cu_3O_{6+y}$ and $HgBa_2CuO_{4+\delta}$---such as the $q_c(\delta)$ relation of Eq. (11) and the validity of the ``universal-dome'' method to determine the hole density $p = p(T_c)$, Sect. XIII. In contrast, superconducting $Bi$-based cuprates with \textit{single} $CuO_2$ plane are more complicated, as they are not only oxygen enriched, but also  \textit{electron doped}. This can be seen in the following diagrams of stepwise ionization of $Bi_2Sr_{2-x}Ln_xCuO_{6}$ ($Ln = La, Bi$). In the \textit{undoped} crystal we have:

\bigskip

\noindent $SrO\;:\;\;\; Sr^{2+} + 2e^- + \;O \rightarrow \; Sr^{2+} + [2e^- + O]\;\;\;\;\;\;\;\;\;\;\;\;\;  \rightarrow  Sr^{2+} + O^{2-}$
\newline
$BiO\;:\;\;\; Bi^{3+} + 3e^- + \;O \rightarrow \; Bi^{3+} + [2e^- + O]\; + \downarrow \overline {e^-\;} | \rightarrow  Bi^{3+} + O^{2-}$
\newline
$CuO_2:\; Cu^{2+} + 2e^- + 2O \rightarrow  Cu^{2+} + [2e^- + O] \;+ \;\;\;O \;\;\rightarrow  Cu^{2+} + 2O^{2-}$
\newline
$BiO\;:\;\;\; Bi^{3+} + 3e^- + \;O \rightarrow \; Bi^{3+} + [2e^- + O]\; + \uparrow \underline {e^-\;} | \rightarrow  Bi^{3+} + O^{2-}$
\newline
$SrO\;:\;\;\; Sr^{2+} + 2e^- + \;O \rightarrow \; Sr^{2+} + [2e^- + O]\;\;\;\;\;\;\;\;\;\;\;\;\;  \rightarrow  Sr^{2+} + O^{2-}$
\medskip

\noindent In the simplest case of {\it doping}, $Ln$ substitutes, in some cells, $Sr$ in both sandwiching layers: 

\medskip 
\noindent $LnO\;:\;\;\; Ln^{3+} + 3e^- + \;O \rightarrow | \overline {e^-} \downarrow + Ln^{3+} + [2e^- + O]\;\;\;\;\;\;\;\;\;\;\;\;\;\;  \rightarrow  Ln^{3+} + O^{2-}$
\newline
$BiO\;:\;\;\; Bi^{3+} + 3e^- + \;O \rightarrow \;\;\;\;\;\;\;\;\;\;\;\;\; Bi^{3+} + [2e^- + O]\;+ \downarrow \overline {e^-\;} | \rightarrow  Bi^{3+} + O^{2-}$
\newline
$CuO_2 :\; Cu^{2+} + 2e^- + 2O \rightarrow  Cu^{2+} \;\;\;\;\;\;\;\;\;\;\;\; + [2e^- + O] + \;\;\;\; O \;\;\rightarrow  \mathbf{Cu} + \;2O^{2-}$
\newline
$BiO\;:\;\;\; Bi^{3+} + 3e^- + \;O \rightarrow \;\;\;\;\;\;\;\;\;\;\;\;\;\; Bi^{3+} + [2e^- + O]\;+ \uparrow \underline {e^-\;} | \rightarrow  Bi^{3+} + O^{2-}$
\newline
$LnO\;:\;\;\; Ln^{3+} + 3e^- + \;O \rightarrow | \underline {e^-} \uparrow +\;\; Ln^{3+} + [2e^- + O]\;\;\;\;\;\;\;\;\;\;\;\;\;  \rightarrow  Ln^{3+} + O^{2-}$
\medskip

\noindent The doped electrons reside pairwise in lattice-site $Cu$ atoms (marked bold). At doping $x=0.5$, an average of half an electron---or a quarter of an electron pair---reaches each unit square of the $CuO_2$ plane. This causes the formation of a $2\times2$ $Cu$ superlattice with afm ordering of magnetic moments, $\mathbf{m}(Cu)$. Because of comparable magnitudes,\cite{1} $|\mathbf{m}(Cu)| \approx |\mathbf{m}(Cu^{2+})|$, they perpetuate the AFM of the host crystal.
There are indications\cite{41,42} that at less doping, $x < 0.5$, some of the excess oxygen enters the $Sr/LnO$ layers, 
\medskip \newline
\noindent $O_{\delta} + 2LnO\rightarrow 2Ln^{3+} + 6e^- + 2O + O_{\delta} \rightarrow \; 2Ln^{3+} + 2[2e^- + O] + 
[2e^- + O_{\delta}] \rightarrow  2Ln^{3+} + 2O^{2-} +O_i^{2-},$

\noindent where each $O_{\delta}$ atom absorbs two electrons from the $Ln$$\rightarrow$$Sr$ substitution and resides as an interstitial $O_i^{2-}$ ion. Accordingly, only a part $\eta = \varphi x$ of the electrons, generated by the $Ln$$\rightarrow$$Sr$ substitution, reaches the $CuO_2$ plane to reduce $Cu^{2+}$ to $Cu$. For $Ln = Bi$, the fraction is $\varphi \simeq 0.5$. It seems energetically favorable that the $2\times2$ $Cu$ superlattice still persists at doping $x<0.5$. The lack of doped electrons from the $Sr/LnO$ layer, necessary to reduce every forth $Cu^{2+}$ ion to $Cu$, is then made up by taking that amount from lattice-site oxygen ions, $O^{2-} - 2e^- \rightarrow O$. This amounts to \textit{hole} doping of the $CuO_2$ plane with density $p=1/4 - \varphi x$.
The $O$ lattice defects form a superlattice of their own. Their magnetic moments, $\mathbf{m}(O)$, give rise to magnetization stripes of incommensurability $q_m(p) \approx p$---an approximation of the derived\cite{1} square-root dependence, Eqs. (3,1), familiar from $La_{2-x}Ae_xCuO_4$. 
As in the latter compounds, the stripes change orientation from diagonal to axes-parallel near $p = 0.056$. 
Table IV shows for $Bi_2Sr_{2-x}Bi_xCuO_{6+\delta}$ the independently determined quantities $(\frac{1}{4} - \frac{x}{2}) \approx p$.

\begin{table}[ht!]
\begin{tabular}{|p{1.cm}|p{1.2cm}|p{1cm}|p{1.1cm}|}
\hline   \hline 
$\;\;\;x$&$\;\frac{1}{4} - \frac{x}{2}$&$\;\;\;p$&$\;\;\;q_m$\\ 
\hline 
$\;0.50$&$\;\;0.00$&$\;0.01$&$\;0.024$\\
$\;0.40$&$\;\;0.05$&$\;0.06$&$\;0.055$\\
$\;0.30$&$\;\;0.10$&$\;0.09$&$\;0.093$\\
$\;0.20$&$\;\;0.15$&$\;0.12$&$\;0.113$\\
\hline   \hline \end{tabular}
\caption{Doping level $x$ of $Bi_2Sr_{2-x}Bi_xCuO_{6+\delta}$, measured with inductively-coupled plasma atomic emission spectroscopy,\cite{41} hole density $p$, measured by Hall effect,\cite{41} and incommensurability of magnetization stripes, $q_m$, measured with neutron scattering.\cite{41}}
\label{table:4}
\end{table}

\section{STRIPES IN OXYGENATED $\mathbf{La_2CuO_{4+\delta}}$ REVISITED}
As mentioned in Sect. III, magnetization stripes with $q_m \simeq 0.115$ parallel to the $Cu$-$O$ bonds are observed\cite{22,27} in oxygen-enriched $La_2CuO_{4+\delta}$, but no accompanying charge-order stripes with $q_c = 2 q_m$, as familiar from $La_{2-x}Ae_xCuO_4$. Instead, charge-order stripes are observed with $q_c = 0.15$ (in quasi-tetragonal r.l.u.) \textit{diagonal} to the $Cu$-$O$ bonds.\cite{21,24,26}
By Eqs. (3, 1), the magnetization stripes with $q_m = 0.115$ are caused by a hole density of $p = 0.12$ in the $CuO_2$ planes. If all of the observed excess oxygen, $y = 0.12$, were to reside ionized at $O(4)$ positions, this would result in a hole density of $p=2\delta=0.24$ in the $CuO_2$ planes. The default raises the possibility that half of the excess oxygen is embedded by a different mechanism---without magnetism but contributing to the observed charge order. 

From the $La_{2-x}Ae_xCuO_4$ compounds it is known that hole doping of the $CuO_2$ planes saturates at a watershed value, $\hat{x} \approx 0.13$, resulting in constant $q_{c,m}(x)$ at higher $Ae$-doping, $x>\hat{x}$ (see Fig. 1).\cite{1} The doped holes then start populating the $LaO$ planes, harbored again pairwise in (lattice-site) $O$ atoms. A similar saturation of doped holes in the $CuO_2$ planes may be at hand in $La_2CuO_{4+\delta}$, starting at $p\simeq 0.12 = 2\delta/2$---except that the $LaO$ planes have already excess oxygen ions, located at $O(4)$ positions. Thus, a different embedding mechanism may commence. From the experience of stripes in oxygen-enriched ($Y$-, $Hg$- or $Bi$-based) cuprates, embedding excess oxygen atoms at ``pore'' sites, $\mathring{O}$, comes to mind. 

Figure 8 shows the $CuO_2$ plane of a model case for $La_2CuO_{4+\delta}$ with both lattice-site $O$ atoms (dark blue) and ozone molecule ions, $\ddot{O}\mathring{O}\ddot{O}$ (turquoise). Each $O$ atom harbors two doped holes that result from excess $O_{\delta/2}^{2-}$ at $O(4)$ positions. Coulomb repulsion spreads the lattice-defect $O$ atoms (with their double holes) to a $4\times4$ superlattice. (If no $\ddot{O}\mathring{O}\ddot{O}$ ions were present, then the $O$ superlattice would give rise to accompanying charge-order stripes parallel to the $Cu$-$O$ bonds with $q_c = 1/4=0.25$.) The $\ddot{O}\mathring{O}\ddot{O}$ ions, formed from interstitial oxygen at pore sites with oxygen neighbors, reside farthest away from the $O$-superlattice sites, that is, at $(a, b) = (2, 2\pm0.5)$ or $(2\pm0.5,2)$ positions near the center of the superlattice 
cells---exemplified by ($2, 1.5)$ in Fig. 8. Magnetization stripes of $q_m = 1/8$ arise from the antiferromagnetic ordering of $\mathbf{m}(O) \ne 0$ moments (red arrows in Fig. 8), whereas the ozone molecule ions have no magnetic contribution, $\mathbf{m}(\ddot{O}\mathring{O}\ddot{O}) = 0$. Approximating the ozone
\medskip

\begin{table}[ht!]
\begin{tabular}{|p{4.2cm}|p{6cm}|p{1.4cm}|p{4.7cm}|}
\hline   \hline 
Quantity of $La_2CuO_{4+\delta}$&Parameters observed in experiment&Ref.&Model parameters in Fig. 8\\ 
\hline 
$\;\;\;\;\;\;\;\;\;\;\;\;\;\;\;\delta$&$\;\;\;\;\;\;\;\;\;\;\;\;\;\;\;\;\;\;\;\;\;\;\;\mathbf{0.12}$&22&$\;\;\;\;\;\;\;\;\;\;\;\;\;\;\;\;\;\;\;0.1325$\\
$\;\;\;\;\;\;\;\;\;\;\;\;\;\;\;2\delta$&$\;\;\;\;\;\;\;\;\;\;\;\;\;\;\;\;\;\;\;\;\;\;\;0.24$&&$\;\;\;\;\;\;\;\;\;\;\;\;\;\;\;\;\;\;\;0.265$\\
\hline 
$\;\;\;\;\;\;\;\;\;\;\;\;\;\;\;q_m(\delta)\;\;$parallel&$\;\;\;\;\;\;\;\;\;\;\;\;\;\;\;\;\;\;\;\;\;\;\;\mathbf{0.115}$&22,27&$\;\;\;\;\;\;\;\;\;\;\;\;\;\;\;\;\;\;\;0.125$\\
$\;\;\;\;\;\;\;\;\;\;\;\;\;\;\;p$&$\;\;\;\;\;\;\;\;\;\;\;\;\;\;\;\;\;\;\;\;\;\;\;0.12$&Eq. (3)&$\;\;\;\;\;\;\;\;\;\;\;\;\;\;\;\;\;\;\;0.14$\\

$\;\;\;\;\;\;\;\;\;\;\;\;\;\;\;2\delta-p$&$\;\;\;\;\;\;\;\;\;\;\;\;\;\;\;\;\;\;\;\;\;\;\;0.12$&&$\;\;\;\;\;\;\;\;\;\;\;\;\;\;\;\;\;\;\;0.125$\\
$\;\;\;\;\;\;\;\;\;\;\;\;\;\;\;\Omega =(2\delta-p)/2$&$\;\;\;\;\;\;\;\;\;\;\;\;\;\;\;\;\;\;\;\;\;\;\;0.06$&&$\;\;\;\;\;\;\;\;\;\;\;\;\;\;\;\;\;\;\;0.0625$\\

\hline
$\;\;\;\;\;\;\;\;\;\;\;\;\;\;\;q_c(\delta)\;\;$diagonal&$\;\;\;\;\;\;\;\;\;\;\;\;\;\;\;\;\;\;\;\;\;\;\;\mathbf{0.15}$&21,24,26&$\;\;\;\;\;\;\;\;\;\;\;\;\;\;\;\;\;\;\;0.177$\\
\hline   \hline \end{tabular}
\caption{Observed parameters of $La_2CuO_{4+\delta}$ (bold) and model parameters of Fig. 8.  The hole density $p$ in the $CuO_2$ planes is calculated from $q_m(\delta)$ with Eqs. (3, 1).  $\Omega$ denotes the density of interstitial oxygen at pore sites, $\mathring{O}$.}
\label{table:5}
\end{table}

\noindent  $(2,2\pm0.5)$ and $(2\pm0.5,2)$ positions by their average, $(2,2)$, yields diagonal charge-order stripes of $q_c = 0.25/\sqrt{2} = 0.177$.  

The model parameters for Fig. 8, chosen for ease of display and listed in Table V, are close to the  parameters $\delta, q_m$ and $q_c$ as experimentally observed in $La_2CuO_{4+\delta}$, although systematically larger. Accordingly, Fig. 8 gives an approximate rendition of the actual $CuO_2$ planes in $La_2CuO_{4+\delta}$. The density of interstitial oxygen at pore sites, $\mathring{O}$, in $La_2CuO_{4.12}$, $\Omega = 0.06$ (Table V) is comparable with that of $YBa_2Cu_3O_{6.67}$ ($\delta = 0.06$, Table II) and $HgBa_2CuO_{4.12}$ ($\delta = 0.06$, Ref. 33). Their transition temperatures are $T_c = 42$ K, 65.5 K,
\medskip 

\includegraphics[width=6in]{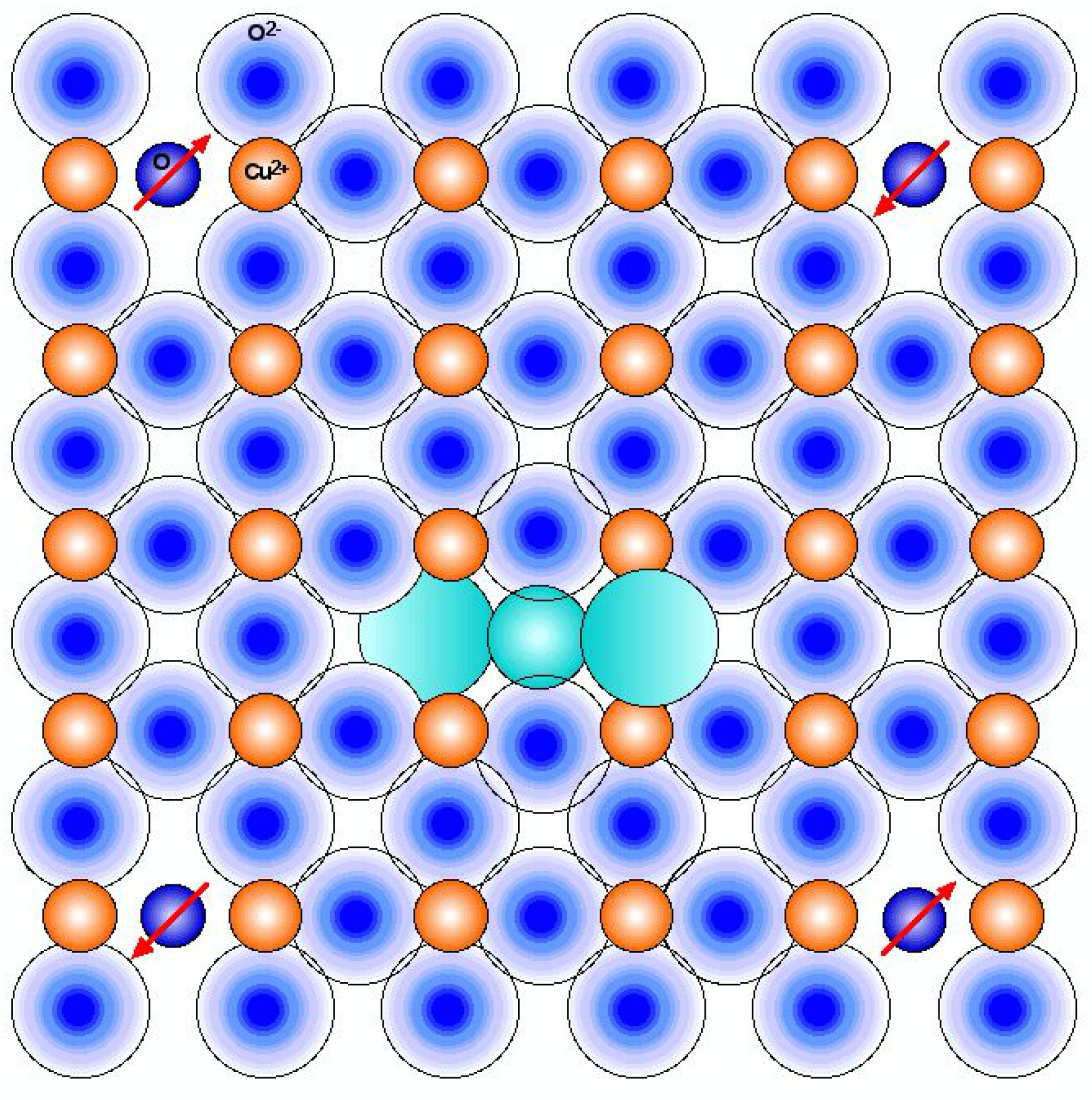} \footnotesize 

\noindent FIG. 8. $CuO_2$ plane of $La_2CuO_{4+\delta}$ with a $4\times4$ superlattice of $O$ atoms (dark blue) and a staggered $4\times4$ superlattice of ozone molecule ions  $\ddot{O}\mathring{O}\ddot{O}$ (turquoise). Model parameters are listed in Table V. Red arrows show the antiferromagnetic ordering of $\mathbf{m}(O)$ moments.
\normalsize 

\noindent   and 55 K, respectively.\cite{28,43,33} 

The finding that oxygen enrichment of $La_2CuO_{4+\delta}$ yields, besides oxygen embedding in oxygen-quartet pores, also hole doping of the $CuO_2$ planes may be a consequence of its staggered $LaO$ double planes with intermediate $O(4)$ sites for $O_{\delta}^{2-}$ ions from oxygen enrichment. Likewise, the hole-doping contribution from \textit{excess} oxygen in $Nd_{2-x}Ce_xCuO_{4+\delta}$ may be a consequence of their staggered $NdO$ double planes. If this argument is valid, the \textit{absence} of $Ln^{3+}O^{2-}$ \textit{double} planes in the $Y$-, $Hg$- and $Bi$-based cuprates would explain their lack of lattice-site $O$ atoms in the $CuO_2$ planes---and accordingly, their lack of square-root dependent stripe incommensurability $q_{c,m}(\delta)$.

\section{HOLE DENSITY IN THE $\mathbf{CuO_2}$ PLANES}

Shortly after the discovery\cite{44} of high-$T_c$ superconductivity in $La_{2-x}Ba_xCuO_4$, various theories were proposed (reviewed in Ref. 45). They all share the view that high-$T_c$ superconductivity arises when the $CuO_2$ planes are doped with holes. Some of the initial theories were discarded quickly because crucial predictions were falsified by experiment. Interest in other initial theories has waned over the decades. Nevertheless, the notion of ``doped holes in the $CuO_2$ planes'' is commonly held. Backed by stoichiometry and charge balance, the notion clearly holds for $La_{2-x}Ae_xCuO_4$, with hole density $p = x$. 
But what is $p$ in $YBa_2Cu_3O_{6+y}$, with its $CuO$ chains, and in the non-stoichiometric, oxygen-enriched cuprates, generated by diffusion and thermal procedures?
Here, a claim of universal $T_c(p)$ dependence has been resorted to.\cite{46} Neglecting the deep chasm in the $T_c(p)$ profile of
$La_{2-x}Ba_xCuO_4$ at $x=1/8$, and the dent in $T_c(p)$ of
$La_{2-x}Sr_xCuO_4$ near $x=1/8$, the profile, also called the ``superconducting dome,'' has been modelled as
\begin{equation}
\frac{T_c(p)}{T_{c,max}} =  1 - \frac{(p - p_{opt}^{SC})^2}{(p_{ons}^{SC} - p_{opt}^{SC})^2} 
= 1 - 82.6  (p - 0.16)^2 \;,
\end{equation}
with maximal transition temperature at optimal doping, $T_c(p_{opt}^{SC}) = T_{c,max}$, and onset of superconductivity, $T_c(p_{ons}^{SC}) = 0$ (see Fig. 9). The onset and optimal doping, $p_{ons}^{SC} \simeq 0.05$ and 
$p_{opt}^{SC} \simeq 0.16$, were obtained from $La_{2-x}Sr_xCuO_4$. This gives the formula on the right of Eq. (14). Inverting the formula furnishes the underdoped {\it (ud)} and overdoped {\it (od)} hole density in the $CuO_2$ plane as
\begin{equation}
p_{ud}^{od}(T_c) = 0.16 \pm \sqrt{\frac{T_{c,max}-T_c}{82.6\;T_{c,max}}} \;,
\end{equation}
in terms of the observed $T_c$ and $T_{c,max}$. Such $p$ values have been widely used to characterize  the doping dependence of stripe incommensurability $q_c(p)$ and of other properties in the pseudogap phase of oxygen-enriched cuprates [later with a more accurate value $p_{ons}^{SC} \simeq 0.056$, which changes the coefficient 82.6 in Eqs.(14, 15) to 91.67].

The universal-dome method---proposed half a decade before charge-order stripes were  discovered\cite{47} in $La_{1.6-x}Nd_{0.4}Sr_xCuO_4$ and (another decade and a half later)\cite{7,8} in  $YBa_2Cu_3O_{6+y}$---assumes that the \textit{same} mechanism of high-$T_c$ superconduction holds both in heterovalent metal-doped, and in oxygen-doped/enriched cuprates. 
There is no \textit{a priori} reason why this should be the case.

What both compound families have in common are $CuO_2$ planes with lattice-defect \textit{oxygen atoms}. Where the two groups differ is the \textit{location} of those atoms---either at anion lattice sites, $\underline{O}$ (besides skirmishing oxygen atoms, $\tilde{O}$), or at interstital sites, $\mathring{O}$, with respective lattice-defect charges $Q^{latt}_{def}(\underline{O}) = +2|e|$ but $Q^{latt}_{def}(\mathring{O})=0$ (relative to the host crystal, see Appendix A) and magnetic moments $|\mathbf{m}(O)| \ne 0$ but $|\mathbf{m}(\ddot{O}\mathring{O}\ddot{O})| = 0$. The lack of electron transfer from $LaO$ to $CuO_2$ planes in the crystal formation of $La_{2-x}Ae_xCuO_4$ (illustrated in Sect. II) can   be regarded as ``hole doping.'' Since the holes reside pairwise in oxygen atoms, we may extend the notion to ``oxygen-atom doping'' of the $CuO_2$ planes, with densities
\begin{equation}
      p(y) = 2\delta(y),
\end{equation}

\includegraphics[width=5.7in]{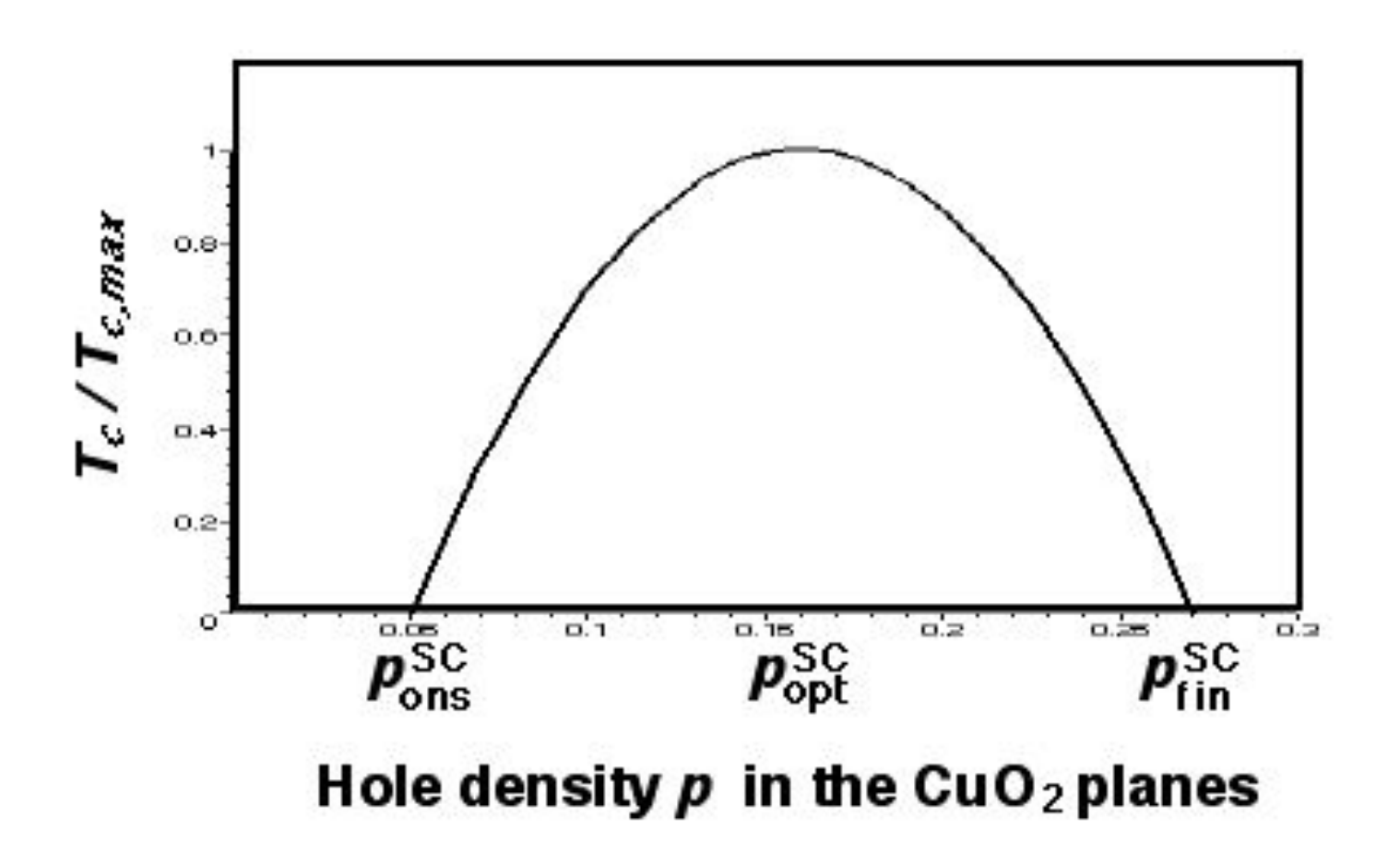} \footnotesize 

\noindent FIG. 9.  Universal dome, $T_c(p)$, of superconducting cuprates, Eq. (14).  \normalsize

\noindent    
respectively. Table II lists values of $2\delta(y)$ for $YBa_2Cu_3O_{6+y}$, obtained with Eq. (11) from stripe data and the use of two family-specific parameters [$\delta(y_c^{ons})$ and $\gamma = -dq_c/d\delta$]. The hole density $p$ can alternatively be obtained from superconducting transition  temperatures with the universal-dome method, Eq. (15). Derived independently, the $2\delta(y)$ and $p(y)$ values in Table II closely agree, satisfying approximately Eq. (16). The agreement has an important implication for the mechanism of high-$T_c$ superconductivity in cuprates, discussed in Sect. XIV. 

Not quite as sure-footed is the universal-dome method for $HgBa_2CuO_{4+\delta}$. Although the onset of superconductivity\cite{33} is also at 2$\delta_{ons}^{SC} =p= 0.05$, the optimal doping is somewhat uncertain, with $\delta_{opt}^{SC} = 0.13 \pm 0.01$ from neutron-scattering experiments but $\delta_{opt}^{SC} = 0.16$ from thermoelectric-power investigations.\cite{33,48} For $Bi$-based cuprates with \textit{two} $CuO_2$ planes per unit cell, such as $Bi_2Sr_2CaCu_2O_{8+\delta}$, the universal-dome method seems quite reliable and is widely used.\cite{49,50,51} 
Conversely, the universal-dome method is not applicable to single-plane $Bi_2Sr_{2-x}Ln_xCuO_{6+\delta}$ ($Ln = La, Bi$), because of both oxygen enrichment and electron doping by $Ln$$\rightarrow$$Sr$ substitution.
In that case, the carrier density is usually obtained from the size of the Fermi surface (Luttinger theorem).\cite{42,52,53}
   

\section{IMPLICATIONS FOR SUPERCONDUCTIVITY}

Although details are not clear, it seems a fairly safe assumption that neutral oxygen atoms in the $CuO_2$ planes affect superconductivity. Just as they give rise to different types of stripes, depending on their position---and position-related properties (lattice-defect charge,  magnetic moment)---they may likewise give rise to different types of superconductivity. A possibility that comes to mind is that in \textit{oxygen-doped/enriched} cuprates, superconducting (Cooper) \textit{pairs} of holes, $2e^+_{\uparrow \downarrow}$, progress by successively moving from interstitial $\mathring{O}$ atoms to neighboring $O^{2-}$ ions, turning them to  $\mathring{O}$ atoms, \textit{etc.},
\begin{equation}
..\;\mathring{O}\;\;O^{2-}O^{2-}O^{2-}.. \rightarrow
..\;O^{2-}\mathring{O}\;\;O^{2-}O^{2-}.. \rightarrow
..\;O^{2-}O^{2-}\mathring{O}\;\;O^{2-}.. \rightarrow
..\;O^{2-}O^{2-}O^{2-}\mathring{O}..
\end{equation}
The spin-singlet configuration ($S=0$) of the hole pair, indicated by the subscript-arrows, and in agreement with the $\mathring{O}$ atom (more accurately, of the $\ddot{O}\mathring{O}\ddot{O}$ ion), gives it a spin-magnetic moment $\mathbf{m}_s(2e^+_{\uparrow \downarrow})=\mathbf{m}_s(\ddot{O}\mathring{O}\ddot{O})=0$.
On the other hand, in \textit{Ae}-doped lanthanum cuprates, 
superconducting pairs of doped holes, $2e^+_{\uparrow \uparrow}$, progress by successively moving from anion-site $O$ atoms to neighboring $O^{2-}$ ions,
\begin{equation}
..\;O\;\;O^{2-}O^{2-}O^{2-}.. \rightarrow
..\;O^{2-}O\;\;O^{2-}O^{2-}.. \rightarrow
..\;O^{2-}O^{2-}O\;\;O^{2-}. \rightarrow
..\;O^{2-}O^{2-}O^{2-}O...
\end{equation}
Here each hole pair, $2e^+_{\uparrow\uparrow}$, has a spin-magnetic moment $\mathbf{m}_s=2\mu_B$ (spin-triplet configuration, $S=1$).  

It turns out that some properties of superconductivity are the same for both processes, (17, 18), but others are different. The universal-dome method obtains the doping of superconductivity onset, $p_{ons}^{SC} = 0.05$, and the optimal doping, $p_{opt}^{SC}=0.16$ of oxygen-doped/enriched cuprates by adopting those of $La_{2-x}Sr_xCuO_4$. 
It is found that the resulting $p$ values agree, for
$YBa_2Cu_3O_{6+y}$, with the $2\delta(y)$ values obtained independently from $YBa_2Cu_3O_{6+y}$ stripe data (Table II).  This confirms that the oxygen-atom doping for onset and maximal superconductivity is the \textit{same} for both compound families, $x_{ons}^{SC} =  p_{ons}^{SC} = 2\delta_{ons}^{SC} = 0.05$ and $x_{opt}^{SC} = p_{opt}^{SC} = 2\delta_{opt}^{SC} = 0.16$.
As far as the onset and maximum of superconductivity are concerned, the exact location of the oxygen atoms in the $CuO_2$ planes doesn't matter:
A threshold density of oxygen atoms is necessary to get superconductivity going, regardless of their exact position. Likewise, maximal \textit{relative} transition temperature $T_c/T_{c,max}$ is achieved at optimal density.

A quantity that is \textit{different} for the processes (17, 18) is the \textit{absolute} transition temperature, $T_c$.
Because of their magnetic moment $|\mathbf{m}_s|=2\mu_B$, the $2e^+_{\uparrow\uparrow}$ hole pairs in process (18) are susceptible to disturbance from magnetic deviations in the crystal, such as magnetization stripes and magnetic impurities. This would lower the transition temperature of compounds with spin-triplet pairs (18), compared to those with spin-singlet pairs (17). The maximal  transition temperatures of comparable compounds with a single $CuO_2$ plane per unit cell,  
listed in Table VI, show clear differences. The temperature $T_{c, max}$ is highest for the oxygen-

\bigskip 
\begin{table}[ht!]
\begin{tabular}{|p{4.cm}|p{0.7cm}|p{1.4cm}|p{1.cm}|p{6.2cm}|}
\hline   \hline 
Compound&$\;\;S$&$T_{c,max}$&Ref.& Comment\\ 
\hline 
$HgBa_2CuO_{4+\delta}$&$\;\;0$&$\;\;95$ K&$\;\;35$&no magnetization stripes\\
$La_{2-x}Sr_xCuO_4$&$\;\;1$&$\;\;36$ K&$\;\;43$&magnetization stripes\\
$La_{1.6-x}Nd_{0.4}Sr_xCuO_4$&$\;\;1$&$\;\;15$ K&$\;\;54$&magn. stripes + $\mathbf{m}(Nd^{3+})$ moments\\
\hline   \hline \end{tabular}
\caption{Maximal transition temperature, $T_{c, max}$, of compounds with a single $CuO_2$ plane per unit cell. $S$ is the spin quantum number of the superconducting hole pair.}
\label{table:6}
\end{table}

\noindent enriched compound ($S=0$) where magnetization stripes are absent. It is much lower for magnetic hole pairs ($S=1$) when subject to magnetization stripes, and still lower when irregularly positioned magnetic moments (from $Nd^{3+}$ with $4f^3$ subshell) are present.

Another superconducting property that is \textit{different} in the two compound families concerns the cooling-curve of X-ray intensity diffracted by charge-order stripes at temperatures \textit{below} $T_c$.
The superconducting hole pairs that don't participate in a supercurrent, \textit{fluctuate} about average positions. In $La_{2-x}Ae_xCuO_4$, Coulomb repulsion keeps $2e^+_{\uparrow\uparrow}$ pairs fluctuating about the $O$ superlattice sites, as can be concluded from the same value of incommensurability, $q_{c,m}(x)$, above and below $T_c$. In contrast, superconducting fluctuations of $2e^+_{\uparrow\downarrow}$ pairs in oxygen-doped/enriched cuprates loosen the axial cohesion between $\ddot{O}\mathring{O}\ddot{O}\mathbf{\ddot{O}}$ motifs. The latter then disperse, causing weaker charge-order stripe-intensity, in thermodynamic proportion of the superconducting/normal states (Gorter-Casimir two-fluid  model of superconductivity). This can be inferred from cooling-curves, shown schematically in Fig. 10. 

Such cooling-curves are distinctly different for $La_{2-x}Ae_xCuO_4$ and $YBa_2Cu_3O_{6+y}$. The signal from $La_{2-x}Ae_xCuO_4$ samples rises continuously with cooling through, and below $T_c$, for $Ae$-doping \textit{below} the watershed value, $x < \hat{x}$. The reason is decreasing thermal agitation, causing increasing coherence length.\cite{1} For $x>\hat{x}$, superconducting fluctuations of $O$ atoms 
to the $LaO$ planes reduces their (average) presence in the $CuO_2$ planes. This reduces the X-ray signal and causes a falling cooling curve below $T_c$.\cite{1} Mixed results of cooling-curves in  \textit{early} experiments on $La_{2-x}Ae_xCuO_4$ have recently been clarified, as discussed in Ref. 1.  
\bigskip 

\includegraphics[width=6in]{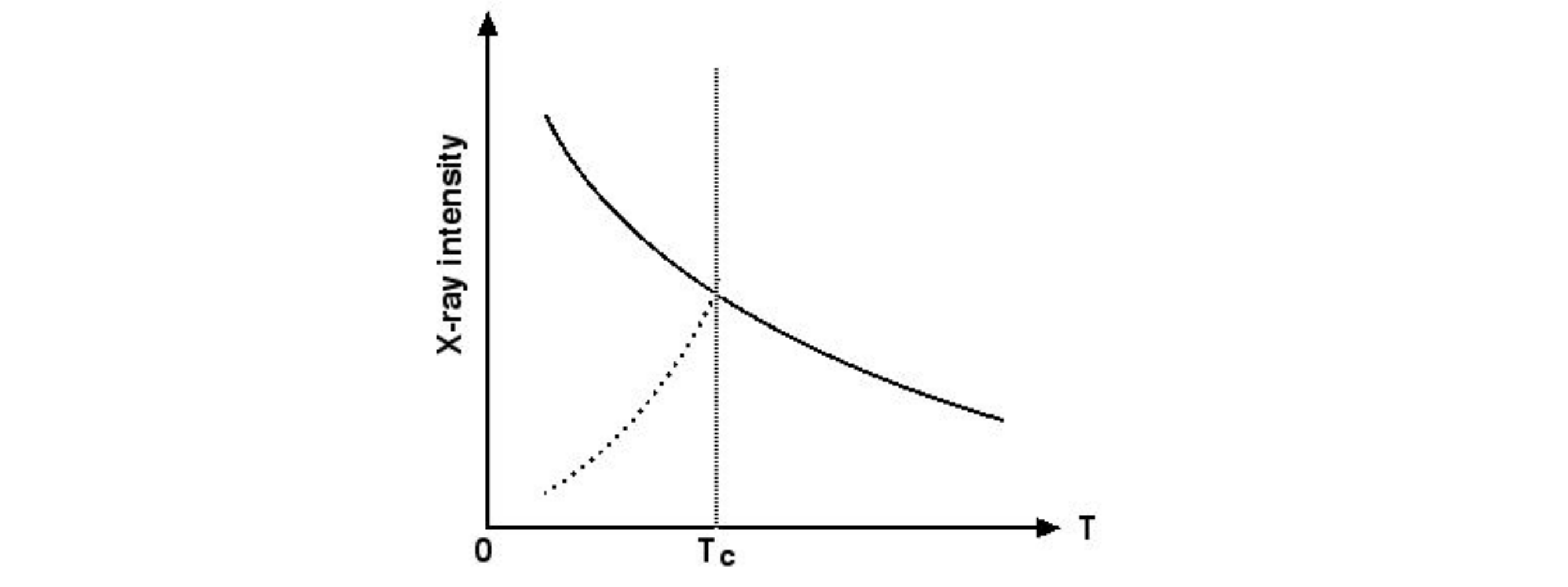} \footnotesize 

\noindent FIG. 10.  Cooling-curves (schematic) of X-ray intensity diffracted by charge-order stripes in the $CuO_2$  planes: Below $T_c$, they are \textit{falling} for $YBa_2Cu_3O_{6+y}$, $Bi_2Sr_2CaCu_2O_{8+\delta}$ and $La_{2-x}Ae_xCuO_4$ ($Ae = Sr, Ba$) if $x>\hat{x}$; but they are \textit{rising} for $La_{2-x}Ae_xCuO_4$ when doped below a watershed value, $x<\hat{x} \approx 0.13$.
  \normalsize 
  
In $YBa_2Cu_3O_{6+y}$ above $T_c$, the cooling-curve also keeps rising upon cooling, due to lesser thermal agitation. However, the curve culminates in a \textit{cusp} at $T_c$. Further cooling leads to a \textit{falling} cooling-curve due to gradual dissolution of the stripe-forming $\ddot{O}\mathring{O}\ddot{O}\mathbf{\ddot{O}}$ trains to isolated $\ddot{O}\mathring{O}\ddot{O}$ molecule ions. From a formal perspective, a falling cooling-curve is conventionally regarded as a competition of (stripe- \textit{vs.} superconducting) order parameters. In the present microscopic view, the falling cooling-curve of $YBa_2Cu_3O_{6+y}$ and $Bi_2Sr_2CaCu_2O_{8+\delta}$ (charge-order ``melting'')\cite{7,8,55} is interpreted as a consequence of quantum fluctuations of $2e^+_{\uparrow\downarrow}$ pairs that loosen the axial cohesion between $\ddot{O}\mathring{O}\ddot{O}\mathbf{\ddot{O}}$ motifs. The same quantum fluctuations also ``soften'' bond-stretching phonons by decreasing, in Fano-like interference, their frequency $\omega(q)$ when the reciprocal-lattice value equals the charge-order incommensurability, $q = q_c$.\cite{55,56}

\section{CONCLUSION}
Charge-order stripes of different types occur when copper oxides are  doped either with heterovalent metal, like $La_{2-x}Sr_xCuO_4$, or oxygen, like $YBa_2Cu_3O_{6+y}$. The difference shows up in the doping dependence of their incommensurability: $q_c(x) \propto \sqrt{x-\check{p}}$ but $q_c(y) \approx 0.3$. The square-root dependence in the former compound family results from Coulomb repulsion between doped holes, residing pairwise in lattice-site $O$ atoms in the $CuO_2$ planes (or between doped electrons, residing pairwise in lattice-site $Cu$ atoms, as in $Nd_{2-x}Ce_xCuO_4$). The almost constant $q_c(y)$ value in the second family---decreasing slightly with more oxygen doping---results from the aggregation of ozone-like molecules, $\ddot{O}\mathring{O}\ddot{O}$, formed from $O^{2-}$ ions of the host with embedded oxygen atoms, $\mathring{O}$, at interstitial sites in the $CuO_2$ planes.
The magnetic moments, $\mathbf{m}(O)$, of the lattice-defect $O$ atoms in the first family arrange antiferromagnetically, which gives rise to accompanying magnetization stripes of incommensurability $q_m(x) = \frac{1}{2} q_c(x)$.
The ozone complex has a vanishing magnetic moment, $\mathbf{m}(\ddot{O}\mathring{O}\ddot{O})=0$, which explains the absence of accompanying magnetization stripes in the second family.
Embedding of excess oxygen atoms in ``pores'' of $CuO_2$ planes is likewise assumed for $HgBa_2CuO_{4+\delta}$ and oxygen-enriched bismuth cuprates.
A combination of characteristics from both families is present in oxygen-enriched $La_2CuO_{4+\delta}$, with both hole doping and ozone complexes, possibly as a consequence of the $LaO$ double layers. 
The validity of the universal-dome method to determine the hole density, $p$, of oxygen-enriched cuprates (not co-doped with electrons) is independently confirmed.
Besides causing different types of stripes, the two types of lattice-defect oxygen, $O$ or $\mathring{O}$, may also cause different types of superconductivity. This could explain the much higher transition temperature $T_{c,max}$ in oxygen-enriched than $Ae$-doped cuprates, as well as the cusped cooling-curves of X-ray intensity diffracted by stripes in the former family.

\appendix
\section{LATTICE-DEFECT CHARGE}
Besides short-range repulsion (due to Pauli's exclusion principle) and van der Waals attraction (from synchronized fluctuating dipoles), long-range \textit{Coulomb} interaction contributes to the lattice energy by the Madelung term,
\begin{equation}
U^{Mad}=\sum_j^{cell} Q(j) V(^{\;\;j}_{site}) \;,
\end{equation}
summed over the atoms of a unit cell. Here $Q(j)$ is the \textit{true} electric charge of atom $j$ and $V(^{\;\;j}_{site})$ is the Madelung potential at the position of $j$. 
We denote a lattice-defect oxygen atom at an anion lattice site in the $CuO_2$ plane by underline, $\underline{O}$. Its contribution to the Madelung energy is
\begin{equation}
\Delta U^{Mad}_{an-sit}(\underline{O})= Q(O) \; V(^{anion}_{\;site}) = 0 \times V(^{anion}_{\;site}) = 0 \;.
\end{equation}
In reference to the corresponding contribution of an $\underline{O}^{2-}$ ion in the (undoped) host crystal,
\begin{equation}
\Delta U^{Mad}_{an-sit}(\underline{O}^{2-})= Q(O^{2-}) \; V(^{anion}_{\;site}) = -2|e| \times V(^{anion}_{\;site}) \;,
\end{equation}
the contribution of the lattice-defect $\underline{O}$ atom, can be expressed as
\begin{equation}
\Delta U^{Mad}_{an-sit}(\underline{O}) \equiv \Delta U^{Mad}_{an-sit}(\underline{O}^{2-}) +Q^{latt}_{def}(\underline{O}) \times V(^{anion}_{\;site}) \;,
\end{equation}
which defines the lattice-defect charge $Q^{latt}_{def}(\underline{O})$.
With Eqs. (A2-4), one obtains $Q^{latt}_{def}(\underline{O})= +2|e|$.
The Madelung potential at an interstitial site is vanishingly small, compared to a lattice site,
\begin{equation}
|V(^{latt}_{site})| \gg |V(^{\;inter}_{stitial})| \approx 0 \;.
\end{equation}
In the limit of $V(^{\;inter}_{stitial}) = 0$,
\textit{any} true charge at an interstitial site, $\mathring{Q}$, gives no contribution to the Madelung energy,
\begin{equation}
\Delta U^{Mad}_{int-sti}(\mathring{Q})= \mathring{Q} \; V(^{\;inter}_{stitial}) = \mathring{Q} \times 0 = 0 \;.
\end{equation}
In this case the lattice-defect charge agrees with the true charge,
$\mathring{Q}^{latt}_{def} = \mathring{Q}$. Specifically for an oxygen atom at a ``pore'' site we have ${Q}^{latt}_{def}(\mathring{O})=0.$

\section{PARAMETERS OF $\mathbf{YBa_2Cu_3O_{6+y}}$ MAGNETIZATION STRIPES}
Squaring Eq. (7) gives, after rearrangement,
\begin{equation}
\frac{16}{g}q_m^2(y)  = y - \check{y}  \; .
\end{equation}
Omitting the magnetization suffix $m$, we label the three oxygen doping levels as $y_i$, $y_j$ and $y_k$ and the corresponding incommensurabilities as $q_i$, $q_j$ and $q_k$. Insertion into Eq. (B1), yields
\begin{equation} \frac{16}{g}q_i^2=y_i-\check{y}\; ,\end{equation}
\begin{equation} \frac{16}{g}q_j^2=y_j-\check{y}\; ,\end{equation}
\begin{equation} \frac{16}{g}q_k^2=y_k-\check{y}\; .\end{equation}
We add Eqs. (B2) and (B3) to obtain, after rearrangement,
\begin{equation}
2\check{y} = y_i + y_j -\frac{16}{g}(q_i^2+q_j^2) \; .
\end{equation}
Substituting $g$ from Eq. (B4) and rearranging yields 
\begin{equation}
\check{y}(2 - \frac{q_i^2+q_j^2}{q_k^2}) = y_i + y_j -y_k \frac{q_i^2+q_j^2}{q_k^2} \; .
\end{equation}
Using the experimental data,\cite{4} $q_i(0.30) = 0.02$, $q_j(0.35) = 0.028$ and $q_k(0.45) = 0.045$, we calculate $\check{y} = 0.27$. 
Insertion of $\check{y}$ into Eq. (B4) gives
\begin{equation}
g = 16\frac{q_k^2}{y_k - \check{y}} \; 
\end{equation}
with a numerical value $g = 0.18$.

\bigskip  \bigskip  
\centerline{ \textbf{ACKNOWLEDGMENTS}}
\noindent I thank Wojciech Tabis and Neven Barišić for information and literature references related to $HgBa_2CuO_{4+\delta}$. Thanks also to Antonio Bianconi for information and literature references on $La_2CuO_{4+\delta}$.
\pagebreak

\end{document}